\documentclass[12pt,draftclsnofoot,onecolumn]{IEEEtran}
\usepackage{cite}
\usepackage{amsmath,amssymb,amsfonts}
\usepackage{algorithm}
\usepackage{algorithmic}
\usepackage{CJK}
\usepackage{longtable,booktabs}
\usepackage{booktabs}
\usepackage{graphicx}
\usepackage{subfigure}
\usepackage{textcomp}
\usepackage{flushend}
\usepackage{amsthm}
\usepackage{upgreek}
\usepackage{graphicx}
\usepackage{supertabular}
\usepackage[bottom]{footmisc}
\usepackage{stfloats}
\usepackage{color}
\usepackage{booktabs}
\usepackage{multirow}
\usepackage{array}
\usepackage{epstopdf}
\UseRawInputEncoding

\def\BibTeX{{\rm B\kern-.05em{\sc i\kern-.025em b}\kern-.08em
    T\kern-.1667em\lower.7ex\hbox{E}\kern-.125emX}}

\newcommand\bfb{\ensuremath{{\mathbf b}}}
\newcommand\bfA{\ensuremath{{\mathbf A}}}
\newcommand\bfg{\ensuremath{{\mathbf g}}}
\newcommand\bfh{\ensuremath{{\mathbf h}}}
\newcommand\bfG{\ensuremath{{\mathbf G}}}
\newcommand\bff{\ensuremath{{\mathbf f}}}
\newcommand\bfH{\ensuremath{{\mathbf H}}}
\newcommand\bfy{\ensuremath{{\mathbf y}}}
\newcommand\bfY{\ensuremath{{\mathbf Y}}}
\newcommand\bfx{\ensuremath{{\mathbf x}}}
\newcommand\bfX{\ensuremath{{\mathbf X}}}
\newcommand\bfz{\ensuremath{{\mathbf z}}}
\newcommand\bfZ{\ensuremath{{\mathbf Z}}}
\newcommand\bfv{\ensuremath{{\mathbf v}}}
\newcommand\bfV{\ensuremath{{\mathbf V}}}
\newcommand\bfa{\ensuremath{{\mathbf a}}}
\newcommand\bfn{\ensuremath{{\mathbf n}}}
\newcommand\bfN{\ensuremath{{\mathbf N}}}
\newcommand\bfT{\ensuremath{{\mathbf T}}}
\newcommand\bfO{\ensuremath{{\mathbf O}}}
\newcommand\rmT{\ensuremath{{\mathrm T}}}
\newcommand\rmt{\ensuremath{{\mathrm t}}}
\newcommand\rmU{\ensuremath{{\mathrm U}}}
\newcommand\rmu{\ensuremath{{\mathrm u}}}
\newcommand\rmB{\ensuremath{{\mathrm B}}}
\newcommand\rmH{\ensuremath{{\mathrm H}}}
\newcommand\rmIB{\ensuremath{{\mathrm {IB}}}}
\newcommand\rmLoS{\ensuremath{{\mathrm {LoS}}}}
\newcommand\rmNLoS{\ensuremath{{\mathrm {NLoS}}}}

\newcommand\rmdiag{\ensuremath{{\mathrm {diag}}}}

\newcommand\rmS{\ensuremath{{\mathrm S}}}
\newcommand\rmI{\ensuremath{{\mathrm I}}}
\newcommand\calT{\ensuremath{{\mathcal T}}}
\newcommand\calO{\ensuremath{{\mathcal O}}}
\newcommand\calN{\ensuremath{{\mathcal N}}}
\newcommand\calC{\ensuremath{{\mathcal C}}}

\newcommand\calM{\ensuremath{{\mathcal M}}}
\newcommand\calA{\ensuremath{{\mathcal A}}}
\newcommand\bbC{\ensuremath{{\mathbb C}}}

\usepackage{pdfpages} 

\begin{document}


\newpage

\title{\textcolor{black}{Deep-Learning} Channel Estimation for IRS-Assisted Integrated Sensing and Communication System}

\author{\IEEEauthorblockN{
Yu Liu,
Ibrahim Al-Nahhal, \textit{Senior Member}, \textit{IEEE}, \\
Octavia A. Dobre, \textit{Fellow}, \textit{IEEE},
and Fanggang Wang, \textit{Senior Member}, \textit{IEEE}}
\thanks{



{Y. Liu and F. Wang are with the State Key Laboratory of Rail Traffic Control and Safety, Frontiers Science Center for Smart High-speed Railway System, Beijing Jiaotong University, Beijing 100044, China. (e-mail: yuliu1@bjtu.edu.cn; wangfg@bjtu.edu.cn).}

{Ibrahim Al-Nahhal and Octavia A. Dobre  are with the Faculty of Engineering and Applied Science, Memorial University, St. John’s, NL A1C 5S7, Canada (e-mail: ioalnahhal@mun.ca; odobre@mun.ca).}

{Digital Object Identifier: 10.1109/TVT.2022.3231727}

{This article is available at: https://ieeexplore.ieee.org/document/9997576}
}
}

\maketitle

\begin{abstract}
  Integrated sensing and communication (ISAC), and intelligent reflecting surface (IRS) are envisioned as revolutionary technologies to enhance spectral and energy efficiencies for next wireless {system} generations.
  For the first time, this paper {focuses} on the channel estimation problem in an IRS-assisted ISAC system.
  This problem is challenging due to the lack of signal processing capacity in passive IRS, as well as the presence of mutual interference between sensing and communication (\textcolor{black}{SAC}) signals in ISAC {systems}.
  A three-stage {approach} is proposed to decouple {the} estimation problem into sub-ones, including the estimation of the direct \textcolor{black}{SAC} channels in {the first} stage, reflected communication channel in {the second} stage, and reflected sensing channel in {the third} stage.
  {The proposed three-stage approach is based on} a deep-learning framework, which involves two different convolutional neural network (CNN) architectures to estimate the channels \textcolor{black}{at the \textcolor{black}{full-duplex} ISAC base station}.
  Furthermore, two types of input-output pairs to train the CNNs are carefully designed\textcolor{black}{, which } \textcolor{black}{affect} the estimation performance under various signal-to-noise ratio conditions and system parameters.
  Simulation results validate the superiority of the proposed estimation approach compared to the \textcolor{black}{least-squares} baseline scheme, and {its} computational complexity is also analyzed.

  \begin{IEEEkeywords}
  Integrated sensing and communication (ISAC), intelligent reflecting surface (IRS), channel estimation, deep-learning, {neural networks}.
  \end{IEEEkeywords}
\end{abstract}

\section{Introduction}\label{sec:intro}
\IEEEPARstart{T}{he} integrated sensing and communication (ISAC) technology has been foreseen as a promising candidate to improve wireless resource utilization and hardware sharing efficiency in {the} {next} wireless {system} {generations} \cite{ref:ISAC-survey,ref:ChModel-b,ref:radar-commun-coexist}.
ISAC merges the sensing and communication (\textcolor{black}{SAC}) functionalities into a single system.
The sensing functionality of ISAC collects and extracts the sensory information from noisy observations.
Thus, it is envisioned as a critical enabler to measure or predict the surrounding wireless environment intelligently.
On the other hand, the communication functionality processes and transfers the received noisy signals, {with the ultimate goal of recovering} the transmitted information accurately. 
Based on these concepts, to successfully integrate {SAC} into a single system and expand ISAC to a wide variety of wireless networks, the performance trade-off between {SAC} has been analyzed in the literature \cite{ref:ISAC-improve-S1,ref:ISAC-improve-S2,ref:ISAC-improve-C1,ref:ISAC-improve-C2,ref:ISAC-improve-C3}.  
The authors in \cite{ref:ISAC-improve-S1} and \cite{ref:ISAC-improve-S2} focused on optimizing the sensing functionality, such as {the} target detection probability and target angle estimation accuracy, while \textcolor{black}{guaranteeing} an acceptable communication performance. 
On the other hand, the sensing-assisted communication scheme that {promotes the} communication performance {was} investigated in \cite{ref:ISAC-improve-C1,ref:ISAC-improve-C2,ref:ISAC-improve-C3}, including the sensing-assisted beam training, tracking, and prediction. 
The aforementioned {works} typically assumed that perfect channel state information (CSI) is {available} {at the receiver side} and rarely considered the channel estimation issue for the ISAC {systems}.
Recently, another interesting study of sensing-assisted communication was {performed} in \cite{ref:ISAC-ChEst-RSU}, which designed the precoder of the roadside units (RSU) and the {received} beamforming of the vehicle. 
With the help of the radar mounted on {the} RSU, the covariance matrices of both {SAC} channels {were} estimated by the echo signals to facilitate the transceiver beamforming design.
However, the {SAC} channels in \cite{ref:ISAC-ChEst-RSU} {were} assumed to share the same dominant paths.
Hence, its estimation approach is only {limited to that} specific {case}.

Intelligent reflecting surface (IRS) has emerged as another promising technique to increase the coverage and capacity of {next} wireless \textcolor{black}{system} {generations}; it enables a programmable wireless propagation environment \cite{ref:IRS-survey,ref:IRS-SCMA-optimize,ref:IRS-SCMA,ref:ChModel-refpower}. 
In general, IRS is an artificial planar surface comprising a large number of low-cost passive reflecting elements.
Each element independently {configures} its phase-shift according to the CSI of the surrounding environment and further controls the reflection of the incident \textcolor{black}{signals} \cite{ref:IRS-survey}.
By properly coordinating the phase-shifts of all the IRS elements, the signal transmission quality and system performance can be {improved}.
This {technology} is referred to as passive beamforming \cite{ref:IRS-beam-gain1,ref:IRS-beam-gain2,ref:IRS-beam-gain3}. 
Note that the above beamforming gain {is based} on \textcolor{black}{an} accurate CSI of the IRS-assisted wireless communication system.
As such, estimating the channels is crucial in such a system.
Since the reflected channel matrix (e.g., user equipment (UE)-IRS-base station (BS) link) {has a} large dimension and does not follow {the} traditional Rayleigh distribution, two challenges come up to the channel estimation \cite{ref:IRS-turn-off} \cite{ref:DL-IRS-ChE-modelDriven}. 
One challenge is the limited estimation accuracy, while the other {one} is the large training overhead.
To overcome {these challenges}, model-driven channel estimation approaches have been widely researched recently, such as the reflection pattern controlled \cite{ref:IRS-ChE-onoff1} \cite{ref:IRS-ChE-onoff2} and element grouping schemes \cite{ref:IRS-ChE-elegroup1} \cite{ref:IRS-ChE-elegroup2}. 
Despite the vital contributions of these works, the above challenges have not been sufficiently addressed yet.
Therefore, the data-driven deep-learning (DL) estimation approaches are further investigated in \cite{ref:DL-IRS-ChE-WCL,ref:IRS-ChE-CNN-group,ref:DL-IRS-ChE-timevary,ref:DL-IRS-ChE-TWC}.
These DL-based approaches reflect the potential to balance the estimation accuracy and training overhead in the IRS-assisted communication systems. 
The mapping between the received signals and channels is successfully characterized by adopting various DL networks, such as convolutional neural network (CNN) \cite{ref:DL-IRS-ChE-WCL,ref:IRS-ChE-CNN-group}, recurrent neural network \cite{ref:DL-IRS-ChE-timevary}, and deep residual learning \cite{ref:DL-IRS-ChE-TWC}.

Since IRS has shown great prospects in enriching communication coverage and spectral/energy efficiency, it is expected to assist the ISAC system in providing better {SAC} performance.
\textcolor{black}{Recently, ISAC and IRS {have been} jointly explored, taking into account their cooperation merits \cite{ref:ChModel-pathloss-SJ,ref:Joint-ISAC-IRS-beam,ref:IRS-ISAC-wavebeam1,ref:IRS-ISAC-wavebeam2}.
Due to the fact that the SAC signals coexist in the IRS-assisted ISAC systems, inherent interference occurs and affects the SAC performance.
Then, the ISAC BS and IRS beamforming, as well as the ISAC waveform, are required to be properly designed for such systems.
The authors in \cite{ref:ChModel-pathloss-SJ} and \cite{ref:Joint-ISAC-IRS-beam} concentrated on the beamforming designs for both the ISAC BS and IRS to provide a trade-off between SAC performance, considering the effect of inherent interference.
On the other hand, the joint ISAC waveform and IRS beamforming designs were investigated for the IRS-assisted ISAC systems in \cite{ref:IRS-ISAC-wavebeam1} and \cite{ref:IRS-ISAC-wavebeam2}, aiming to mitigate the inherent interference while improving the communication performance under the constraints of sensing metrics.
It is worth noting that an accurate CSI is necessary for all the {aforementioned} designs.
The channel estimation problem in such IRS-assisted ISAC systems is challenging due to the inherent interference; to the best of the authors' knowledge, {this problem} has not been investigated yet.} 

Considering the research gap in the existing literature, this paper proposes a novel three-stage channel estimation approach for an IRS-assisted ISAC multiple-input single-output (MISO) system.
Each \textcolor{black}{stage} is devised at the \textcolor{black}{full-duplex (FD)} ISAC BS, and then a DL estimation framework is developed correspondingly, along with input-output pairs designs. 
In particular, the contributions {of this paper} are summarized as follows:
\begin{enumerate}[]
  \item A three-stage estimation approach is proposed to estimate the \textcolor{black}{SAC} channels {of} the IRS-assisted ISAC system.
      It decouples the overall estimation problem into sub-ones {by controlling an on/off state of the IRS or BS transmission. The proposed estimator} successively {estimates} the direct \textcolor{black}{SAC}, reflected UE-IRS-BS, and reflected BS-target-IRS-BS channels.
      The proposed approach is the first attempt to provide a practical channel estimation for such a system and successfully solves the estimation difficulty caused by the {inherent} interference.
  \item The pilot transmission protocol for each stage is precisely designed to facilitate the \textcolor{black}{SAC} channels estimation.
      It involves the design of the pilot sequences adopted at the {FD} ISAC BS, pilot sequences employed at the UE, and IRS phase-shift vectors.
  \item A CNN-based DL framework is developed at the ISAC BS to estimate the \textcolor{black}{SAC} channels in three stages.
      Considering the different propagation environments of the direct and reflected channels, two CNN architectures are carefully designed to form {the proposed DL} framework.
      {The first} CNN is devised for the direct \textcolor{black}{SAC} channels estimation in the first stage, while the other one is employed in the second and third stages to estimate the reflected channels.
  \item Two types of input-output pairs are \textcolor{black}{proposed} for the CNNs.
      The first type of input-output pair is built on the original received \textcolor{black}{SAC} signals, while the second one relies on the least-squares (LS) estimates of the \textcolor{black}{SAC} channels.
      Inspired by the data augmentation in data analysis, the training dataset is further enriched to enhance the estimation {accuracy}.
  \item The computational complexity of the proposed approach for the input generation and CNN-based online estimation is analyzed.
      Their mathematical formulations are deduced in terms of the required number of real additions and multiplications.
      Quantitative simulations indicate that the complexity of the proposed approach is acceptable {compared to the LS baseline estimator}.
  \item Extensive simulations {are performed to} assess the estimation performance of the proposed approach.
      Numerical results reveal that the proposed approach achieves considerable enhancement in the estimation accuracy compared to the {LS} baseline scheme under various signal-to-noise ratio (SNR) conditions and system parameters.
      It also possesses promising generalization capacity under a wide range of SNR regions.
\end{enumerate}

The remainder of this paper is organized as follows:
Sections \ref{sec:system} and \ref{sec:DL-CE-appraoch} introduce the system model and proposed three-stage channel estimation approach for the IRS-assisted ISAC MISO system, respectively.
The complexity analysis and simulation results are presented in Sections \ref{sec:complexity} and \ref{sec:simulation}, respectively.
Finally, the conclusions are drawn in Section \ref{sec:conclusion}.

{\textit{Notations}}: Boldface lowercase and uppercase letters {represent} vectors and matrices, respectively.
$\mathcal{CN}(\mu,\sigma^2)$ \textcolor{black}{is} the \textcolor{black}{probability density function} of a random variable following the complex Gaussian distribution with mean $\mu$ and variance $\sigma^2$. 
For any vector $\bfx$, $\text{diag}\{\bfx\}$ returns a diagonal square matrix whose diagonal consists of the elements of $\bfx$.
The operators $(\cdot)^{\rm H}$, $(\cdot)^\mathrm{T}$, $(\cdot)^*$, $(\cdot)^{-1}$, \textcolor{black}{$(\cdot)^\dag$}, $\lfloor \cdot \rfloor$, vec$[\cdot]$, $\Re\{\cdot\}$, $\Im\{\cdot\}$, $\mathbb E\{\cdot\}$, and $\|\cdot\|_F$ stand for the Hermitian, transpose, conjugate, inverse, \textcolor{black}{pseudoinverse}, floor, vectorization, real part, imaginary part, expectation, and \textcolor{black}{Frobenius norm} of their arguments, respectively.
$\jmath=\sqrt{-1}$ denotes the imaginary unit.
$\calN_a^b=\{a,a+1,\ldots,b\}$ denotes the index set from integer $a$ to $b$, and $a<b$.

\section{System Model}\label{sec:system}
\begin{figure}
\centering
\includegraphics[width=3in]{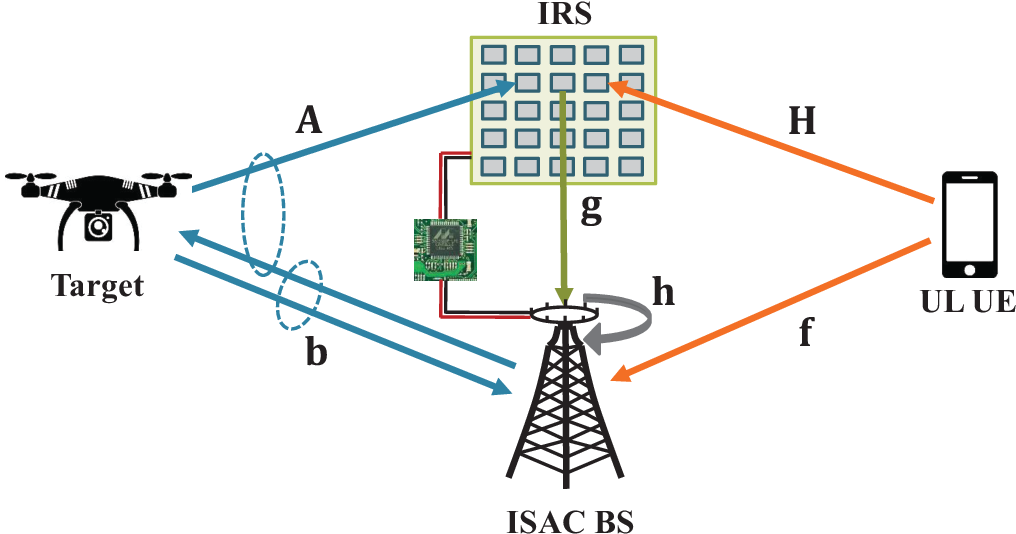}
\caption{IRS-assisted ISAC MISO system model.}
\label{fig:System}
\end{figure}

Consider an IRS-assisted ISAC MISO system with \textcolor{black}{an FD} ISAC BS, a target, an uplink (UL) UE, and IRS, as illustrated in Fig. \ref{fig:System}.
The ISAC BS is equipped with $M$ transmit antennas and one \textcolor{black}{receive} antenna to simultaneously sense the target and communicate with the UL UE \textcolor{black}{that} has $M$ transmit antennas.
IRS composed of $L$ passive reflecting elements is employed to assist the \textcolor{black}{SAC}.
\textcolor{black}{Note that the IRS location impacts system performance (e.g., achievable rate).
Here, the IRS is chosen to be near the ISAC BS to \textcolor{black}{ensure} the maximum benefit of the system performance \cite{ref:ChModel-refpower}.}
For target sensing, the \textcolor{black}{FD} ISAC BS receives the echo signals from the direct BS-target-BS and reflected BS-target-IRS-BS channels.
Let $\bfb\in\bbC^{M \times 1}$, $\bfA\in\bbC^{M\times L}$, and $\bfg\in\bbC^{1 \times L}$ denote the sensing channel coefficients of the BS-target-BS, BS-target-IRS, and IRS-BS links, respectively.
The UL communication signals are received at the ISAC BS through the direct UE-BS and reflected UE-IRS-BS channels.
Define $\bff\in\bbC^{1\times M}$ and $\bfH\in\bbC^{L\times M}$ as the communication channel coefficients of the UE-BS and UE-IRS links, respectively.
The self-interference (SI) is introduced to the ISAC BS from the SI channel, $\bfh\in\bbC^{M\times 1}$, due to the FD mode.

\begin{figure}
\centering
\includegraphics[width=3.5in]{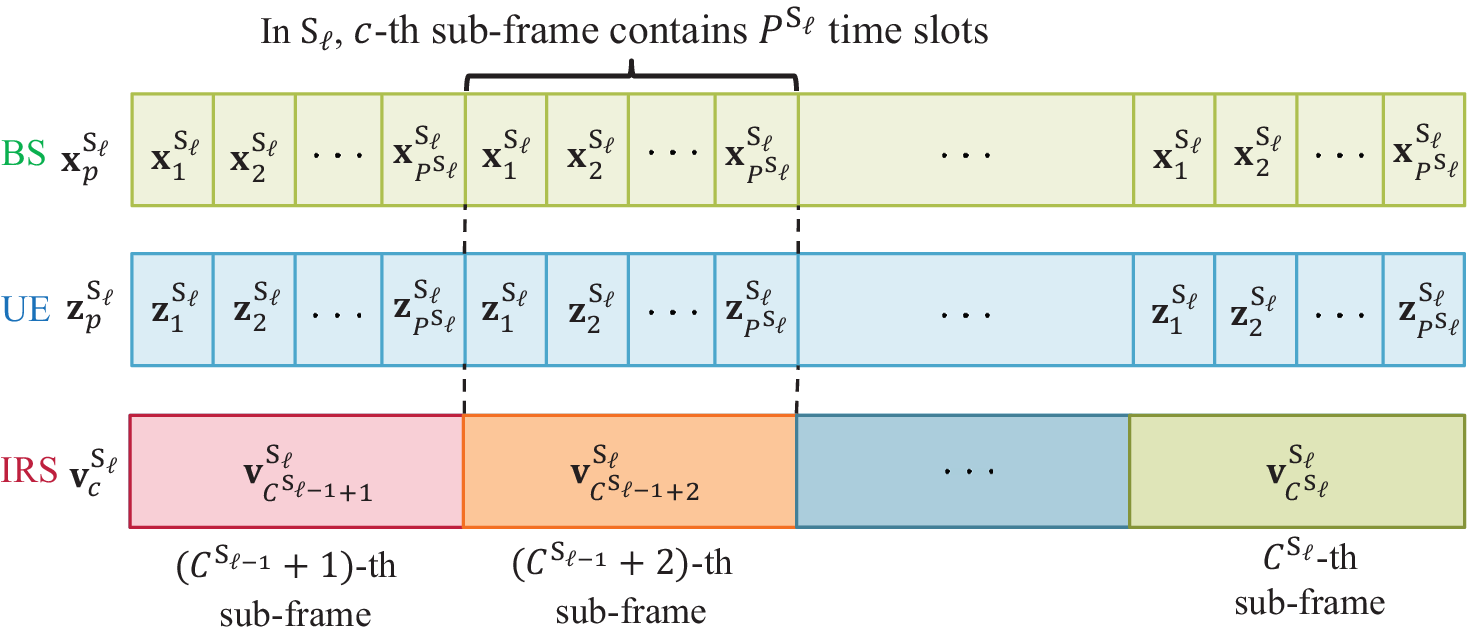}
\caption{Pilot transmission protocol.}
\label{fig:Pilot}
\end{figure}

A pilot transmission protocol is \textcolor{black}{designed} and illustrated in Fig. \ref{fig:Pilot} to estimate the \textcolor{black}{SAC} channels of the IRS-assisted ISAC MISO system.
Regarding the proposed three-stage approach, define $\rmS_\ell$ as the $\ell$-th estimation stage, $\ell\in\calN_1^3$.
As shown in Fig. \ref{fig:Pilot}, \textcolor{black}{in $\rmS_\ell$}, the ISAC BS and UL UE respectively send pilot sequences in \textcolor{black}{$C^{\rmS_\ell}-C^{\rmS_{\ell-1}}$} sub-frames {with} the $c$-th, $c \in\calN_{C^{\rmS_{\ell-1}}+1}^{C^{\rmS_\ell}}$, sub-frame \textcolor{black}{consisting} of {$P^{\rmS_\ell}$} time slots.
Note that the duration of the $\ell$-th estimation stage is from the $(C^{\rmS_{\ell-1}}+1)$-th to the $C^{\rmS_\ell}$-th sub-frame \textcolor{black}{with} $C^{\rmS_0}=1$.
\textcolor{black}{At the $p$-th, $p\in\calN_1^{P^{\rmS_\ell}}$, time slot in each sub-frame}, the pilot signal vectors adopted at the ISAC BS and UL UE are defined as \textcolor{black}{$\bfx_{p}^{\rmS_\ell}\in\bbC^{M\times1}$ and $\bfz_{p}^{\rmS_\ell}\in\bbC^{M\times1}$}, respectively.
\textcolor{black}{Their corresponding pilot signal matrices in one sub-frame are written as $\bfX^{\rmS_\ell} = [\bfx_1^{\rmS_\ell}, \bfx_2^{\rmS_\ell}, \ldots, \bfx_{P^{\rmS_\ell}}^{\rmS_\ell}]\in\bbC^{M\times P^{\rmS_\ell}}$ and $\bfZ^{\rmS_\ell} = [\bfz_1^{\rmS_\ell},\bfz_2^{\rmS_\ell}, $ $ \ldots, \bfz_{P^{\rmS_\ell}}^{\rmS_\ell}]\in\bbC^{M\times P^{\rmS_\ell}}$, respectively.}
Moreover, denote the IRS phase-shift vector in the $c$-th sub-frame by \textcolor{black}{$\bfv_c^{\rmS_\ell}\in\bbC^{1\times L}$}.
\textcolor{black}{Here,} \textcolor{black}{$\bfv_c^{\rmS_\ell}$} remains constant within one sub-frame.
Thus, the received signal at the $p$-th time slot in the $c$-th sub-frame, \textcolor{black}{$y_{c,p}^{\rmS_\ell}$}, at the ISAC BS can be expressed as
\begin{align}
y_{c,p}^{\rmS_\ell}& = \underbrace{ (\bfb^\rmH + {\bfg}\, \rmdiag\{\bfv_c^{\rmS_\ell}\} \bfA^\rmH)\bfx_{p}^{\rmS_\ell} }_{\text {Sensing signal}}
 + \underbrace{ \bfh^\rmH\bfx_{p}^{\rmS_\ell} }_{\text {Residual SI}} \notag \\
& \quad\quad + \underbrace{(\bff + {\bfg}\, \rmdiag\{\bfv_c^{\rmS_\ell}\} \bfH)\bfz_{p}^{\rmS_\ell}}_{\text {UL communication signal}} + n_{c,p}^{\rmS_\ell}, 
\label{eq:y_cp}
\end{align}
where $\bfv_c^{\rmS_\ell}=[\beta_c e^{\jmath\varphi_{c,1}}, \beta_c e^{\jmath\varphi_{c,2}}, \ldots, $ $\beta_c e^{\jmath\varphi_{c,L}}]$, in which $\beta_c\in[0,1]$ and $\varphi_{c,l}\in[0,2\pi)$ with $l\in\calN_1^L$ are the amplitude and phase-shift of the $l$-th IRS element, respectively.
The {additive white Gaussian} noise $n_{c,p}^{\rmS_\ell}$ follows $\mathcal{CN}(0,\sigma^2)$ with zero-mean and variance $\sigma^2$. 
\textcolor{black}{It is worth mentioning that the propagation environment between the transmit and receive antennas of the ISAC BS tends to be slow-varying \cite{ref:SI-estimation1, ref:SI-estimation3, ref:SI-estimation4, ref:SI-estimation5}. 
As such, the SI channel, $\bfh$, can be pre-estimated by an LS estimator at the ISAC BS, and the residual SI term in (1) is compensated before estimating the remaining channels \cite{ref:SI-estimation1}.}
As seen from \eqref{eq:y_cp}, it is apparent that \textcolor{black}{${\bfg}\, \rmdiag\{\bfv_c^{\rmS_\ell}\}= \bfv_c^{\rmS_\ell}\, \rmdiag\{\bfg\}$}.
The equivalent reflected channels of \textcolor{black}{the} BS-target-IRS-BS and UE-IRS-BS links are given by $\bfG_\rmt = \bfA\, \rmdiag\{{\bfg^\rmH}\}\in\bbC^{M\times L}$ and $\bfG_\rmu = \rmdiag\{\bfg\}\, \bfH \in\bbC^{L\times M}$, respectively. 
Therefore, \textcolor{black}{$y_{c,p}^{\rmS_\ell}$} in \eqref{eq:y_cp} can be reformulated as
\begin{align}
y_{c,p}^{\rmS_\ell} = \underbrace{(\bfb^\rmH + \bfv_c^{\rmS_\ell} {\bfG_\rmt}^\rmH)\bfx_{p}^{\rmS_\ell}}_{\text {Sensing signal}} + \underbrace{(\bff + \bfv_c^{\rmS_\ell} \bfG_\rmu)\bfz_{p}^{\rmS_\ell}}_{\text {UL communication signal}} + n_{c,p}^{\rmS_\ell}.
\label{eq:y_cp_re}
\end{align}
Note that the received \textcolor{black}{SAC} signals interfere with each other and are hard to be decoupled at the ISAC BS.
Hence, the channels in \eqref{eq:y_cp_re} are \textcolor{black}{challenging} to be estimated. 
In the following, {the} aim is to estimate the \textcolor{black}{SAC} channels {(i.e., $\bfb$, $\bfG_\rmt$, $\bff$, and $\bfG_\rmu$)} for {the} IRS-assisted ISAC system based on the designed pilot transmission protocol{, as shown in Fig. \ref{fig:Pilot}}.

\section{Proposed Three-Stage Estimation Approach}\label{sec:DL-CE-appraoch}
In this section, a DL-based three-stage channel estimation approach is proposed.
Firstly, each of the estimation stages is \textcolor{black}{explained in detail.}
Then, the input-output pairs for the DL networks are designed \textcolor{black}{for} the three stages.
In light of these, a CNN-based DL estimation framework is further developed.

\subsection{Design \textcolor{black}{of} Three-Stage \textcolor{black}{Approach}}
To simplify the estimation problem in \eqref{eq:y_cp_re}, a three-stage channel estimation \textcolor{black}{approach} is proposed here.
In {$\rmS_1$}, the direct \textcolor{black}{SAC} channels {(i.e., $\bfb$ and $\bff\}$)} are estimated {when} all the IRS elements are turned off by the BS backhaul link \cite{ref:IRS-turn-off}.
By turning on the IRS and controlling the on/off state of {the} BS transmission in {$\rmS_2$ and $\rmS_3$}, the reflected channels of the UE-IRS-BS and BS-target-IRS-BS links {(i.e., $\bfG_\rmu$ and $\bfG_\rmt$)} are successively estimated.
The details of each stage are introduced as follows.

\subsubsection{First Stage}
{From} the first to the $C^{\rmS_1}$-th sub-frames, all the IRS elements are turned off to estimate the direct \textcolor{black}{SAC} channels \textcolor{black}{(i.e., $\bfb$ and $\bff$)}.
Based on the \textcolor{black}{proposed} pilot transmission protocol in Fig. \ref{fig:Pilot}, the received signal in the $c$-th sub-frames, $\bfy_c^{\rmS_1}\in\bbC^{1\times P^{\rmS_1}}$, at the ISAC BS can be expressed as
\begin{align}
\bfy_c^{\rmS_1} = \bfb^\rmH \bfX^{\rmS_1} + \bff \bfZ^{\rmS_1} + \bfn_c^{\rmS_1}, \quad c\in\calN_1^{C^{\rmS_1}},
\label{eq:y_S1}
\end{align}
\textcolor{black}{where $\bfn_c^{\rmS_\ell} = [n_{c,1},n_{c,2},\ldots,n_{c,P^{\rmS_\ell}}]\in\bbC^{1\times P^{\rmS_\ell}}$ is the noise vector in $\rmS_\ell$.
Here, $\ell=1$ and $\rmS_\ell=\rmS_1$.}
To avoid the impact of interference between \textcolor{black}{SAC} signals on the estimation accuracy, let {$\bfX^{\rmS_1}$} and {$\bfZ^{\rmS_1}$} be orthogonal.
Denote a matrix by {${\bf R}=[(\bfX^{\rmS_1})^\rmT, (\bfZ^{\rmS_1})^\rmT]^\rmT\in\bbC^{2M\times P^{\rmS_1}}$} with $P^{\rmS_1}\geqslant2M$ \cite{ref:Bayesian-pilot-num}, and design it as a discrete Fourier transform (DFT) matrix.
The $(q,w)$-th entry in ${\bf R}$ is written as ${\bf R}_{(q,w)}=\frac 1 {\sqrt{M}} e^{\jmath\frac{2\pi}{P^{\rmS_1}}qw}$.
Considering the low pilot overhead demand, let $P^{\rmS_1}=2M$.
\textcolor{black}{Consequently}, the pilot matrix \textcolor{black}{$\bfX^{\rmS_1}$} is constructed by selecting the first to the $M$-th rows of ${\bf R}$, whereas \textcolor{black}{$\bfZ^{\rmS_1}$} is generated by the rest of \textcolor{black}{${\bf R}$ (i.e., from the $(M+1)$-th to the $2M$-th rows)}.

\subsubsection{Second Stage}
To estimate the reflected communication channel $\bfG_\rmu$ in \textcolor{black}{$\rmS_2$}, the transmission of ISAC BS is turned off, while all the IRS elements are turned on during the \textcolor{black}{$(C^{\rmS_1}+1)$-th to the $C^{\rmS_2}$-th} sub-frames.
The received UL communication signal in the $c$-th sub-frame, $\bfy_c^{\rmS_2}\in\bbC^{1\times P^{\rmS_2}}$, at the ISAC BS is given by
\begin{align}
\bfy_c^{\rmS_2} = \Big(\bff + \bfv_c^{\rmS_2} \bfG_\rmu\Big)\bfZ^{\rmS_2} + \bfn_c^{\rmS_2}, \quad c\in\calN_{C^{\rmS_1}+1}^{C^{\rmS_2}}.
\label{eq:y_S2}
\end{align}
\textcolor{black}{The pilot signal matrix, {$\bfZ^{\rmS_2}$}, adopted in {$\rmS_2$} is designed as an $M\times P^{\rmS_2}$ DFT matrix with {$P^{\rmS_2}\geq M$}, and its $(q,w)$-th entry is formulated as ${\bf Z}_{(q,w)}^{\rmS_2}=\frac 1 {\sqrt{M}} e^{\jmath\frac{2\pi}{P^{\rmS_2}}qw}$.}
Taking into account the demand for low-pilot overhead, let \textcolor{black}{$P^{\rmS_2}=M$}. 
The IRS phase-shift matrix from the $(C^{\rmS_1}+1)$-th to the $C^{\rmS_2}$-th sub-frames is denoted by $\bfV^{\rmS_2}=[(\bfv_{C^{\rmS_1}+1}^{\rmS_2})^\rmT,  $ $ (\bfv_{C^{\rmS_1}+2}^{\rmS_2})^\rmT,\ldots,(\bfv_{C^{\rmS_2}}^{\rmS_2})^\rmT]^\rmT \in\bbC^{(C^{\rmS_2}-C^{\rmS_1})\times L}$.
Refer to \cite{ref:DL-IRS-ChE-TWC}, $\bfV^{\rmS_2}$ is devised as a DFT matrix with $C^{\rmS_2}-C^{\rmS_1} \geq L$.
This is an optimal scheme to enhance the received signal power at ISAC BS and guarantee \textcolor{black}{accurate} channel estimation \cite{ref:DL-IRS-ChE-TWC}.

\subsubsection{Third Stage}
To estimate the reflected sensing channel $\bfG_\rmt$ in \textcolor{black}{$\rmS_3$}, the ISAC BS transmission and IRS are simultaneously turned on.
Within the $(C^{\rmS_2}+1)$-th to the $C^{\rmS_3}$-th sub-frames, \textcolor{black}{the} ISAC BS receives both \textcolor{black}{SAC} signals from the direct and reflected channels.
The received signal in the $c$-th sub-frame, $\bfy_c^{\rmS_3}\in\bbC^{1\times P^{\rmS_3}}$, at the ISAC BS is
\begin{align}
& \bfy_c^{\rmS_3} = \Big(\bfb^\rmH + \bfv_c^{\rmS_3} {\bfG_\rmt}^\rmH\Big)\bfX^{\rmS_3}  \notag \\
& \quad\quad + \Big(\bff + \bfv_c^{\rmS_3} \bfG_\rmu\Big)\bfZ^{\rmS_3} + \bfn_c^{\rmS_3}, \quad c\in\calN_{C^{\rmS_2}+1}^{C^{\rmS_3}}.
\label{eq:y_S3}
\end{align}
\textcolor{black}{In $\rmS_3$, the adopted pilot matrices (i.e., $\bfX^{\rmS_3}$ and $\bfZ^{\rmS_3}$) are devised as the $M\times P^{\rmS_3}$ DFT matrices with $P^{\rmS_3}=M$, {considering the low-pilot overhead demand}.
The $(q,w)$-th entry in them is given by ${\bf X}_{(q,w)}^{\rmS_3}={\bf Z}_{(q,w)}^{\rmS_3}=\frac 1 {\sqrt{M}} e^{\jmath\frac{2\pi}{P^{\rmS_3}}qw}$.}
Denote the IRS phase-shift matrix in \textcolor{black}{$\rmS_3$} by $\bfV^{\rmS_3}=[(\bfv_{C^{\rmS_2}+1}^{\rmS_3})^\rmT, $ $ (\bfv_{C^{\rmS_2}+2}^{\rmS_3})^\rmT,\ldots,(\bfv_{C^{\rmS_3}}^{\rmS_3})^\rmT]^\rmT \in\bbC^{(C^{\rmS_3}-C^{\rmS_2})\times L}$ and design it as a DFT matrix with $C^{\rmS_3}-C^{\rmS_2} \geq L$ \cite{ref:DL-IRS-ChE-TWC}.

\subsection{Input-Output Pairs Design} \label{sec:ChEst_IO}
Based on the received signals at the ISAC BS, the generation of input-output pairs for the DL networks is designed in the three estimation stages.

\subsubsection{First Stage} \label{sec:ChEst_IO_S1}
To estimate the direct \textcolor{black}{SAC} channels {(i.e., $\bfb$ and $\bff$)}, two types of input-output pairs for the DL are devised in $\rmS_1$.
{Let $\rmI_k$ denote the $k$-th input-output pair type, $k\in\calN_1^2$.}
Then, the first input of the DL, $\bfT^{\rmS_1\rmI_1}$, is directly generated by the received \textcolor{black}{SAC} signals in \eqref{eq:y_S1} as
\begin{align}
\bfT^{\rmS_1\rmI_1} = \Big[\Re\Big\{\big[\bfy_1^{\rmS_1}, \ldots, \bfy_{C^{\rmS_1}}^{\rmS_1}\big]\Big\}, \Im\Big\{\big[\bfy_1^{\rmS_1}, \ldots, \bfy_{C^{\rmS_1}}^{\rmS_1}\big]\Big\}\Big]^{\rmT}.
\label{eq:I_S1I1}
\end{align}

The second input of the DL, $\bfT^{\rmS_1\rmI_2}$, is constructed by utilizing the LS estimation results of the direct \textcolor{black}{SAC} channels {that} \textcolor{black}{are} respectively denoted by $\bar\bfb\in\bbC^{M \times 1}$ and $\bar\bff\in\bbC^{1 \times M}$.
Since the pilot matrices \textcolor{black}{(i.e., $\bfX^{\rmS_1}$ and $\bfZ^{\rmS_1}$)} are orthogonal, the estimated $\bar\bfb$ and $\bar\bff$ are respectively given by
\begin{align}
\bar\bfb = \mathbb{E}\big\{ (\bfy_c^{\rmS_1}(\bfX^{\rmS_1})^\dag)^\rmH \big\} = \bfb + \mathbb{E}\big\{(\bar{\bfn}_c^{\rmS_1})^\rmH\big\},
\label{eq:bar_b}
\end{align}
and
\begin{align}
\bar\bff = \mathbb{E}\big\{ \bfy_c^{\rmS_1}{(\bfZ^{\rmS_1}})^\dag \big\} = \bff + \mathbb{E}\big\{ \tilde{\bfn}_c^{\rmS_1} \big\},
\label{eq:bar_f}
\end{align}
where \textcolor{black}{$(\bfX^{\rmS_1})^\dag=(\bfX^{\rmS_1})^\rmH (\bfX^{\rmS_1}(\bfX^{\rmS_1})^\rmH)^{-1}$,  $(\bfZ^{\rmS_1})^\dag=(\bfZ^{\rmS_1})^\rmH (\bfZ^{\rmS_1}(\bfZ^{\rmS_1})^\rmH)^{-1}$, $\bar{\bfn}_c^{\rmS_1}={\bfn}_c^{\rmS_1}(\bfX^{\rmS_1})^\dag$ and $\tilde{\bfn}_c^{\rmS_1}={\bfn}_c^{\rmS_1}(\bfZ^{\rmS_1})^\dag$.}
Based on \eqref{eq:bar_b} and \eqref{eq:bar_f}, $\bfT^{\rmS_1\rmI_2}$ is given by
\begin{align}
\bfT^{\rmS_1\rmI_2} = \Big[\Re\Big\{\big[\bar\bfb^\rmT, \bar\bff\big]\Big\}, \Im\Big\{\big[\bar\bfb^\rmT, \bar\bff\big]\Big\} \Big]^\rmT.
\label{eq:I_S1I2}
\end{align}

For both inputs $\bfT^{\rmS_1\rmI_1}$ and $\bfT^{\rmS_1\rmI_2}$, the corresponding output of the DL {in $\rmS_1$}, $\bfO^{\rmS_1}$, is constructed by the ground truth of the direct \textcolor{black}{SAC} channels {(i.e., $\bfb$ and $\bff$)} as
\begin{align}
\bfO^{\rmS_1} = \Big[\Re\Big\{\big[\bfb^\rmT, \bff\big]\Big\}, \Im\Big\{\big[\bfb^\rmT, \bff\big]\Big\} \Big]^\rmT.
\label{eq:O_S1}
\end{align}

\subsubsection{Second Stage}\label{sec:ChEst_IO_S2}
To estimate the reflected communication channel $\bfG_\rmu$, two types of input-output pairs for the DL are designed \textcolor{black}{in $\rmS_2$}.
Note that the direct communication channel has been estimated as $\hat\bff\in\bbC^{1\times M}$ after \textcolor{black}{$\rmS_1$}.
Thus, by utilizing the estimated $\hat\bff$ and the received communication signals in \eqref{eq:y_S2}, the first type of input for the DL, $\bfT^{\rmS_2 \rmI_1}$, is generated as
\begin{align}
& \bfT^{\rmS_2 \rmI_1} = \Big[\Re\Big\{\big[\bfy_{C^{\rmS_1}+1}^{\rmS_2}, \bfy_{C^{\rmS_1}+2}^{\rmS_2} \ldots, \bfy_{C^{\rmS_2}}^{\rmS_2}, \hat{\bff}\big]\Big\}, \notag\\
& \quad\quad\Im\Big\{\big[\bfy_{C^{\rmS_1}+1}^{\rmS_2}, \bfy_{C^{\rmS_1}+2}^{\rmS_2} \ldots, \bfy_{C^{\rmS_2}}^{\rmS_2}, \hat{\bff}\big]\Big\}\Big]^{\rmT}. 
\label{eq:I_S2I1}
\end{align}

The second input of the DL, $\bfT^{\rmS_2 \rmI_2}$, is constructed by the LS estimate of the reflected communication channel, defined as $\bar{\bfG}_\rmu\in\bbC^{L\times M}$.
To obtain $\bar{\bfG}_\rmu$, the rough estimation of the UL reflected signal is firstly derived by
\begin{align}
\tilde{\bfy}_c^{\rmS_2} & = \bfy_c^{\rmS_2} - \hat{\bff}\bfZ^{\rmS_2} \notag \\
& = \bfv_c^{\rmS_2} \bfG_\rmu \bfZ^{\rmS_2} + \tilde{\bfn}_c^{\rmS_2},
\quad c\in\calN_{C^{\rmS_1}+1}^{C^{\rmS_2}},
\label{eq:ytilde_cS2}
\end{align}
where $\tilde{\bfn}_c^{\rmS_2}=(\bff-\hat{\bff})\bfZ^{\rmS_2} + {\bfn}_c^{\rmS_2}$.
By separating the orthogonal pilot matrix \textcolor{black}{$\bfZ^{\rmS_2}$} from $\tilde{\bfy}_c^{\rmS_2}$ in \eqref{eq:ytilde_cS2}, the \textcolor{black}{resulted} $\bar{\bfy}_c^{\rmS_2}\in\bbC^{1\times M}$ is formulated as
\begin{align}
\bar{\bfy}_c^{\rmS_2} = \tilde{\bfy}_c^{\rmS_2} (\bfZ^{\rmS_2})^\dag = \bfv_c^{\rmS_2} \bfG_\rmu + \bar{\bfn}_c^{\rmS_2}, \quad c\in\calN_{C^{\rmS_1}+1}^{C^{\rmS_2}},
\label{eq:ybar_cS2}
\end{align}
where $\bar{\bfn}_c^{\rmS_2}=\tilde{\bfn}_c^{\rmS_2}{(\bfZ^{\rmS_2})}^\dag$.
Then, from the $(C^{\rmS_1}+1)$-th to the $C^{\rmS_2}$-th sub-frames, the matrix form of \eqref{eq:ybar_cS2} is expressed as
\begin{align}
\bar{\bfY}^{\rmS_2} = \bfV^{\rmS_2} \bfG_\rmu + \bar{\bfN}^{\rmS_2},
\end{align}
where $\bar{\bfY}^{\rmS_2} = [(\bar{\bfy}^{\rmS_2}_{C^{\rmS_1}+1})^\rmT, (\bar{\bfy}^{\rmS_2}_{C^{\rmS_1}+2})^\rmT, \ldots, (\bar{\bfy}^{\rmS_2}_{C^{\rmS_2}})^\rmT]^\rmT$ and $\bar{\bfN}^{\rmS_2} = [(\bar{\bfn}^{\rmS_2}_{C^{\rmS_1}+1})^\rmT, (\bar{\bfn}^{\rmS_2}_{C^{\rmS_1}+2})^\rmT, \ldots, (\bar{\bfn}^{\rmS_2}_{C^{\rmS_2}})^\rmT]^\rmT$. 
Accordingly, the LS estimate of the reflected communication channel is
\begin{align}
\bar{\bfG}_\rmu = \big(\bfV^{\rmS_2}\big)^\dag \bar{\bfY}^{\rmS_2}.
\label{eq:bar_Gu}
\end{align}
Based on \eqref{eq:bar_Gu}, $\bfT^{\rmS_2 \rmI_2}$ is constructed by
\begin{align}
\bfT^{\rmS_2 \rmI_2} = \Big[\Re\{{\rm{vec}}[\bar{\bfG}_\rmu]\}, \Im\{{\rm{vec}}[\bar{\bfG}_\rmu]\}\Big]^\rmT.
\label{eq:I_S2I4}
\end{align}

For these two types of inputs, the corresponding output of the DL \textcolor{black}{in $\rmS_2$}, $\bfO^{\rmS_2}$, is generated by the ground truth of the reflected communication channel $\bfG_\rmu$ as
\begin{align}
\bfO^{\rmS_2} = \Big[\Re\{{\rm{vec}}[\bfG_\rmu]\}, \Im\{{\rm{vec}}[\bfG_\rmu]\}\Big]^\rmT.
\label{eq:O_S2}
\end{align}

\subsubsection{Third Stage}\label{sec:ChEst_IO_S3}
Similar to the previous stages, two types of input-output pairs for the DL are designed to estimate the reflected sensing channel $\bfG_\rmt$ \textcolor{black}{in $\rmS_3$}.
Based on the estimated channels \textcolor{black}{(i.e., $\hat\bfb$, $\hat\bff$, and $\hat\bfG_\rmu$)} from the previous stages and the received signals in \eqref{eq:y_S3}, the first input of the DL, $\bfT^{\rmS_3 \rmI_1}$, is designed as
\begin{align}
& \bfT^{\rmS_3 \rmI_1} = \Big[ \Re\Big\{ \big[\bfy_{C^{\rmS_2}+1}^{\rmS_3}, \bfy_{C^{\rmS_2}+2}^{\rmS_3} \ldots, \bfy_{C^{\rmS_3}}^{\rmS_3}, \hat{\bff}, \hat{\bfb}^\rmT, {\rm {vec}}\{\hat{\bfG}_\rmu\} \big] \Big\}, \notag\\
& \quad\quad \Im\Big\{ \big[\bfy_{C^{\rmS_2}+1}^{\rmS_3}, \bfy_{C^{\rmS_2}+2}^{\rmS_3} \ldots, \bfy_{C^{\rmS_3}}^{\rmS_3}, \hat{\bff}, \hat{\bfb}^\rmT, {\rm {vec}}\{\hat{\bfG}_\rmu\} \big] \Big\} \Big]^{\rmT}. 
\label{eq:I_S3I1}
\end{align}

The second input of the DL, $\bfT^{\rmS_3 \rmI_2}$, is generated by the LS estimate of the reflected sensing channel\textcolor{black}{,} denoted by $\bar{\bfG}_\rmt\in\bbC^{M\times L}$.
The rough estimation of the reflected sensing signal is firstly given by
\begin{align}
\tilde{\bfy}_c^{\rmS_3} & = \bfy_c^{\rmS_3} - \big(\hat{\bff} + \bfv_c^{\rmS_3} \hat{\bfG}_\rmu\big)\bfZ^{\rmS_3} - \hat{\bfb}^\rmH\bfX^{\rmS_3} \notag \\
& = \bfv_c^{\rmS_3} \bfG_\rmt^\rmH \bfX^{\rmS_3} + \tilde{\bfn}_c^{\rmS_3}, \quad c\in\calN_{C^{\rmS_2}+1}^{C^{\rmS_3}},
\label{eq:ytilde_cS3}
\end{align}
where $\tilde{\bfn}_c^{\rmS_3} = \big(\bff-\hat{\bff} + \bfv_c^{\rmS_3}\big(\bfG_\rmu - \hat{\bfG}_\rmu\big)\big)\bfZ^{\rmS_3} + \big(\bfb-\hat{\bfb}\big)\bfX^{\rmS_3} + {\bfn}_c^{\rmS_3}$.
As in \textcolor{black}{$\rmS_2$}, to obtain $\bar{\bfG}_\rmt$, the orthogonal pilot matrix $\bfX^{\rmS_3}$ is separated from $\tilde{\bfy}_c^{\rmS_3}$ and the derived $\bar{\bfy}_c^{\rmS_3}$ is
\begin{align}
\bar{\bfy}_c^{\rmS_3} = \tilde{\bfy}_c^{\rmS_3} {(\bfX^{\rmS_3})}^\dag = \bfv_c^{\rmS_3} \bfG_\rmt^\rmH + \bar{\bfn}_c^{\rmS_3}, \quad c\in\calN_{C^{\rmS_2}+1}^{C^{\rmS_3}},
\label{eq:ybar_cS3}
\end{align}
where $\bar{\bfn}_c^{\rmS_3}=\tilde{\bfn}_c^{\rmS_3}{(\bfX^{\rmS_3})}^\dag$.
Then, the matrix form of \textcolor{black}{\eqref{eq:ybar_cS3}} is written as
\begin{align}
\bar{\bfY}^{\rmS_3} = \bfV^{\rmS_3} \bfG_\rmt^\rmH + \bar{\bfN}^{\rmS_3},
\end{align}
where $\bar{\bfY}^{\rmS_3}$ and $\bar{\bfN}^{\rmS_3}$ are defined similarly as $\bar{\bfY}^{\rmS_2}$ and $\bar{\bfN}^{\rmS_2}$, respectively. 
Therefore, the LS estimate of the reflected sensing channel is given by
\begin{align}
\bar{\bfG}_\rmt = \big(\bar{\bfY}^{\rmS_3}\big)^\rmH \big({\bfV}^{\rmS_3}\big)^\dag.
\label{eq:bar_Gt}
\end{align}
With $\bar{\bfG}_\rmt$ in \eqref{eq:bar_Gt}, $\bfT^{\rmS_3 \rmI_2}$ is generated as
\begin{align}
\bfT^{\rmS_3 \rmI_2} = \Big[\Re\{{\rm{vec}}[\bar{\bfG}_\rmt]\}, \Im\{{\rm{vec}}[\bar{\bfG}_\rmt]\}\Big]^\rmT.
\label{eq:I_S3I4}
\end{align}

Correspondingly, the output of the DL \textcolor{black}{in $\rmS_3$}, $\bfO^{\rmS_3}$, is constructed by the ground truth of the reflected sensing channel $\bfG_\rmt$ as
\begin{align}
\bfO^{\rmS_3} = \Big[\Re\{{\rm{vec}}[\bfG_\rmt]\}, \Im\{{\rm{vec}}[\bfG_\rmt]\}\Big]^\rmT. 
\label{eq:O_S3}
\end{align}

\subsection{Proposed CNN-\textcolor{black}{Based} DL Framework}
\label{sec:CNN}

\begin{figure}
\centering
\subfigure[ ]
{\begin{minipage}[b]{0.48\textwidth}
\includegraphics[width=1\textwidth]{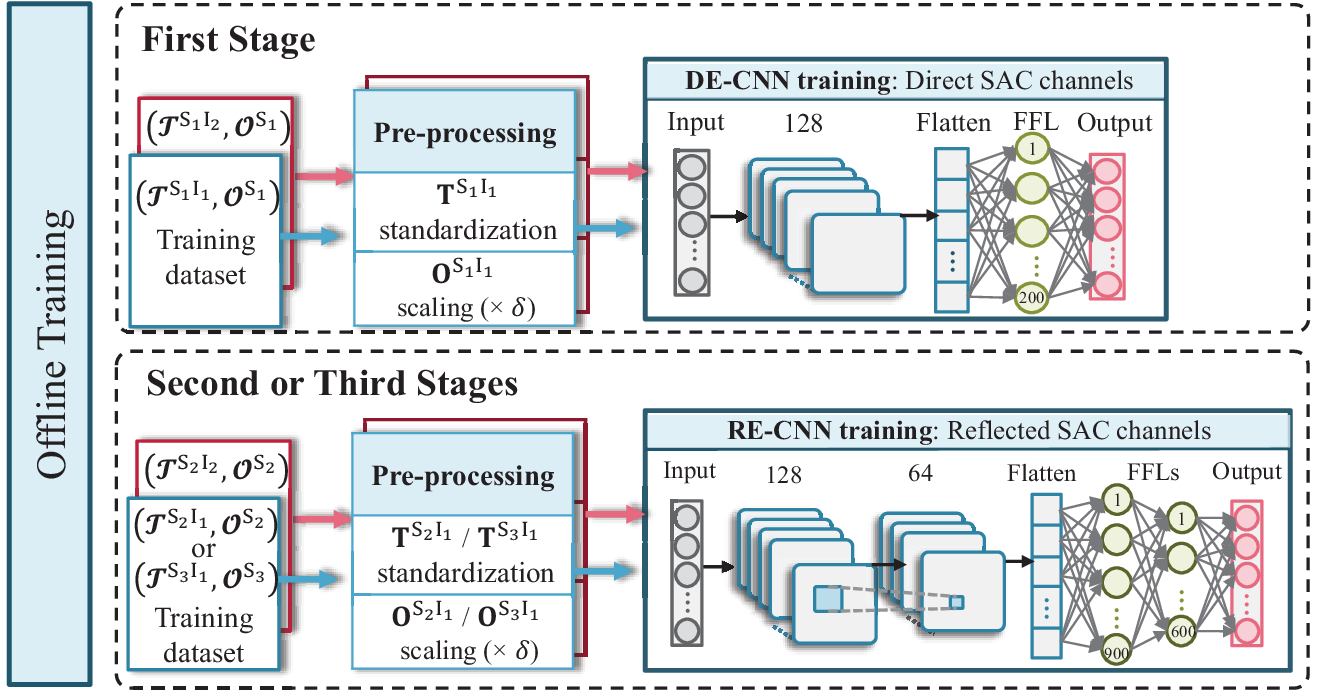}
\end{minipage}}
\subfigure[ ]
{\begin{minipage}[b]{0.48\textwidth}
\includegraphics[width=1\textwidth]{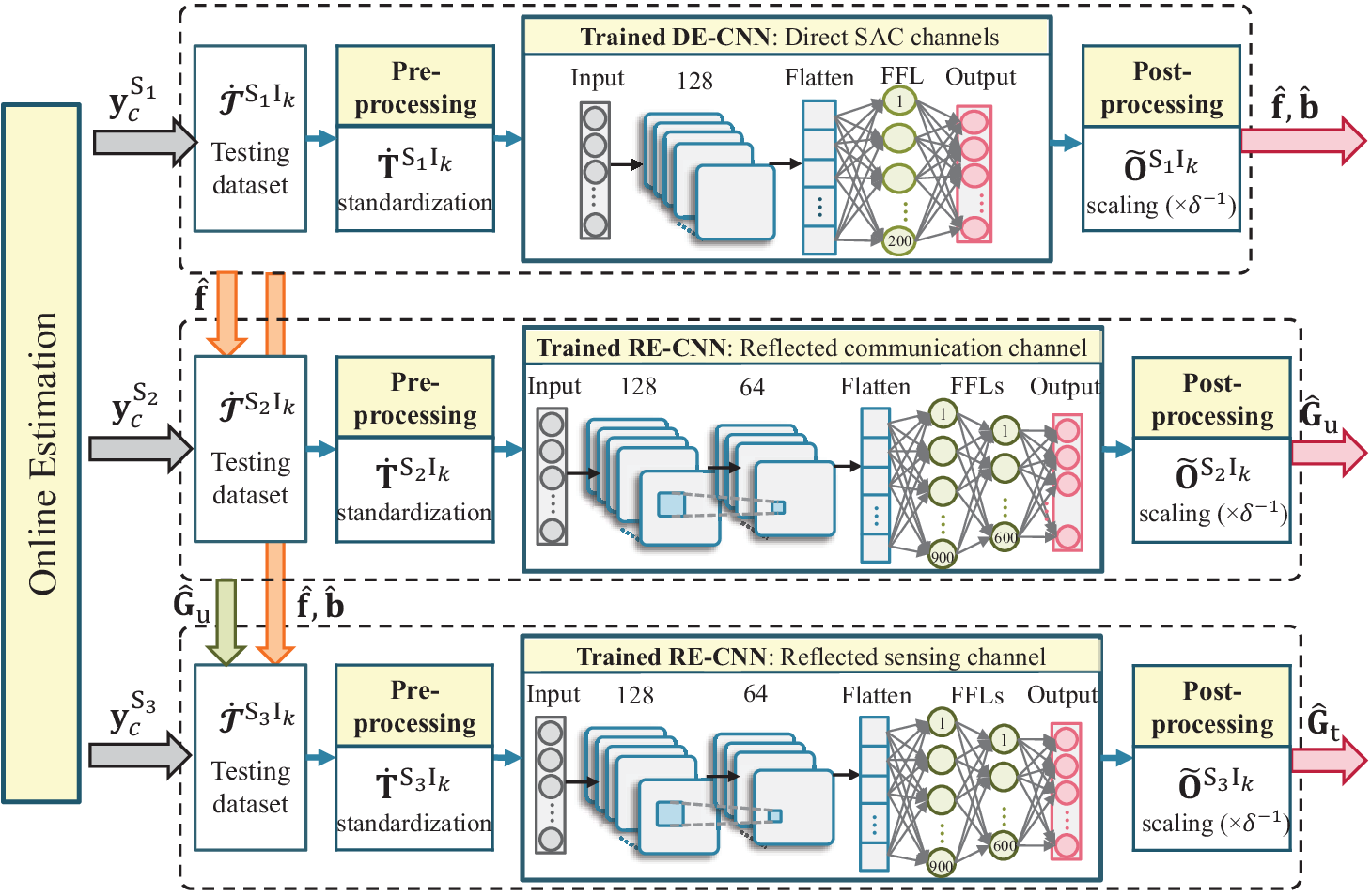}
\end{minipage}}
\caption{The proposed CNN-based DL framework: (a) Offline training, (b) Online estimation.}
\label{fig:CNN_Diag}
\end{figure}

On the basis of the proposed three-stage estimation approach and the designed input-output pairs, a CNN-based DL framework is proposed.
It comprises the offline training and online estimation phases, as depicted in Fig. \ref{fig:CNN_Diag}.
The training dataset is generated to train the proposed DL framework.


\subsubsection{\textcolor{black}{Training Dataset Generation}}\label{sec:ChEst_training_set}
For the three estimation stages, the training datasets of the DL are generated by the devised input-output pairs.
Regarding the $k$-th type of input-output pair {in $\rmS_\ell$}, \textcolor{black}{the corresponding training dataset} is denoted by
\begin{align}
& (\calT^{\rmS_\ell\rmI_k},\calO^{\rmS_\ell})=\Big\{ \big(\bfT^{\rmS_\ell\rmI_k}_{(1,1)},\bfO^{\rmS_\ell}_{(1)}\big), \big(\bfT^{\rmS_\ell\rmI_k}_{(1,2)},\bfO^{\rmS_\ell}_{(1)}\big),\ldots,  \big(\bfT^{\rmS_\ell\rmI_k}_{(1,U)},\bfO^{\rmS_\ell}_{(1)}\big), \notag\\
& \qquad
\big(\bfT^{\rmS_\ell\rmI_k}_{(2,1)},\bfO^{\rmS_\ell}_{(2)}\big), \ldots,  \big(\bfT^{\rmS_\ell\rmI_k}_{(V,U)},\bfO^{\rmS_\ell}_{(V)}\big)  \Big\}, \quad \ell\in\calN_1^3, \quad k\in\calN_1^2,
\label{eq:I_S1_set}
\end{align}
where $\big(\bfT^{\rmS_\ell\rmI_k}_{(v,u)},\bfO^{\rmS_\ell}_{(v)}\big)$ \textcolor{black}{represents} the $(v,u)$-th, $v\in\calN_{1}^V$, $u\in\calN_{1}^U$, training sample.
For further understanding, take the training dataset $(\calT^{\rmS_1\rmI_1},\calO^{\rmS_1})$ as an example to illustrate its generation process in detail.
Inspired by the data augmentation in data analysis, the inputs of the DL in this dataset are constructed by using $V$ received signals \textcolor{black}{in $\rmS_1$} (i.e., \textcolor{black}{SAC} signals formulated in \eqref{eq:y_S1}), and $U-1$ copies of the $v$-th, $v\in\calN_1^V$, are generated by adding {synthetic additive white Gaussian noise} to the original \textcolor{black}{SAC} channels with $\text{SNR}_\text{ch} =\frac{{\cal P}_\text{ch} }{\sigma_\text{ch}^2}$.
\textcolor{black}{The synthetic noise follows ${\cal CN}(0,\sigma_\text{ch}^2)$ with zero-mean and variance $\sigma_\text{ch}^2$, and the power of the original channel is ${\cal P}_\text{ch}$ \cite{ref:DL-IRS-ChE-WCL}}.
Accordingly, {the} noise corrupted \textcolor{black}{SAC} channels can be utilized to obtain the {signal copies} and further generate the copy samples (i.e., $\big(\bfT^{\rmS_1\rmI_1}_{(v,u)},\bfO^{\rmS_1}_{(v)}\big)$, $v\in\calN_{1}^V$, $u\in\calN_{2}^U$).
In conjunction with their original versions (i.e., $\big(\bfT^{\rmS_1\rmI_1}_{(v,1)},\bfO^{\rmS_1}_{(v)}\big)$, $v\in\calN_{1}^V$), the training dataset $(\calT^{\rmS_1\rmI_1},\calO^{\rmS_1})$ is generated.
Note that the generation of these copies intends to enrich the training dataset and improve the estimation performance of the \textcolor{black}{proposed} DL.
Moreover, the training datasets for \textcolor{black}{$\rmS_2$ and $\rmS_3$} are constructed similarly as \textcolor{black}{in $\rmS_1$}.

\subsubsection{Offline Training}
As seen from Fig. \ref{fig:CNN_Diag}(a), the offline training phase for each stage contains two procedures, including the data pre-processing and the DL network training.
The data pre-processing is performed on the training datasets, \textcolor{black}{consisting} of the input standardization and the output scaling with a factor of $\delta=10^4$.
By utilizing the pre-processed samples, the networks in the proposed framework are trained to obtain well-trained networks.
Two different CNN architectures are devised to realize this framework.
One CNN is employed in \textcolor{black}{$\rmS_1$} for the direct \textcolor{black}{SAC} channels estimation, referred to as direct estimation CNN (DE-CNN), while the other {CNN} is adopted in {$\rmS_2$ and $\rmS_3$} for the reflected \textcolor{black}{SAC} channels estimation, namely reflected estimation CNN (RE-CNN).

The detailed offline training phase is illustrated here.
Given the training dataset of the $k$-th type in \textcolor{black}{$\rmS_\ell$} (i.e., $(\calT^{\rmS_\ell\rmI_k},\calO^{\rmS_\ell})$), the $(v,u)$-th input-output pair in this dataset is pre-processed as $\big(\bar{\bfT}^{\rmS_\ell\rmI_k}_{(v,u)},\bar{\bfO}^{\rmS_\ell}_{(v)}\big)$.
Then, during the training procedure, the designed CNN approximates $\bar{\bfO}^{\rmS_\ell}_{(v)}$ as
\begin{align}
\bar{\bfO}^{\rmS_\ell}_{(v)} \approx f^{\rmS_\ell\rmI_k} \Big(\bar{\bfT}^{\rmS_\ell\rmI_k}_{(v,u)}; \Theta^{\rmS_\ell\rmI_k} \Big), \quad v\in\calN_1^V, \quad u\in\calN_1^U,
\end{align}
where $f^{\rmS_\ell\rmI_k}(\cdot \,; \Theta^{\rmS_\ell\rmI_k})$ represents the CNN function that corresponds to the $k$-th type input-output pair in \textcolor{black}{$\rmS_\ell$}, and $\Theta^{\rmS_\ell\rmI_k}$ denotes all \textcolor{black}{CNN} hyperparameters. 
To update $\Theta^{\rmS_\ell\rmI_k}$, the designed CNN minimizes the mean square error (MSE) of the loss function, ${\cal L}^{\rmS_\ell\rmI_k}$, as
\begin{align}
{\cal L}^{\rmS_\ell\rmI_k} = \frac 1 {VU} \sum_{v\in\calN_1^{V}, u\in\calN_1^{U}} \bigg(f^{\rmS_\ell\rmI_k}\Big(\bar{\bfT}_{(v,u)}^{\rmS_\ell\rmI_k}; \Theta^{\rmS_\ell\rmI_k}\Big)-\bar{\bfO}_{(v)}^{\rmS_\ell}\bigg)^2.
\label{eq:loss}
\end{align}

\begin{table}[t] 
\newcommand{\tabincell}[2]{\begin{tabular}{@{}#1@{}}#2\end{tabular}}
\centering
\caption{HyperParameters of DE-CNN and RE-CNN\textcolor{black}{.}} \label{table:CNN}
\begin{tabular}{p{1cm}<{\centering}| p{2cm}<{\centering}| p{5cm}<{\centering}| p{2.5cm}<{\centering}| p{2.8cm}<{\centering}}
\hline
                        & {\bf Layer type}   & {\bf Tensor size} & {\bf Kernel Size}& {\bf Activation function} \\
\hline
\hline
\multirow{5}{*}{\bf {DE-CNN}} & \multirow{2}{*}{Input}  & $\rmS_1\rmI_1$: $4MC^{\rmS_1}$ & \multirow{2}{*}{-} & \multirow{2}{*}{-} \\
                        &                    & $\rmS_1\rmI_2$: $4M\quad\,\,\,\,$ &                     &  \\ \cline{2-5}
                        &  CL                & $128$                         &$4\times 1$          & {\it tanh}   \\ \cline{2-5}
                        &  FFL               & $200$                         &-                    & {\it linear} \\ \cline{2-5}
                        &  Output            & $\rmS_1$: $4M$                &-                    & -      \\ \cline{1-5}
\multirow{10}{*}{\bf {RE-CNN}} & \multirow{4}{*}{Input} & $\rmS_2\rmI_1$: $2M(C^{\rmS_2}-C^{\rmS_1}+1)\quad\,\,\,\,$ & \multirow{4}{*}{-} & \multirow{4}{*}{-} \\
                        &             & $\rmS_2\rmI_2$: $2ML\qquad\qquad\qquad\qquad\,\,$     &            &    \\
                        &             & $\rmS_3\rmI_1$: $2M(C^{\rmS_3}-C^{\rmS_2}+L+2)$                  &            &    \\
                        &             & $\rmS_3\rmI_2$: $2ML\qquad\qquad\qquad\qquad\,\,\,\,$     &            &    \\ \cline{2-5}
                        &  CL         & $128$                                            &$4\times 1$ & {\it tanh} \\ \cline{2-5}
                        &  CL         & $64$             &$4\times 1$   & {\it tanh}   \\ \cline{2-5}
                        &  FFL        & $600$            &-             & {\it linear} \\ \cline{2-5}
                        &  FFL        & $900$            &-             & {\it linear} \\ \cline{2-5}
                        &  \multirow{2}{*}{Output}       & $\rmS_2$: $2ML$  &\multirow{2}{*}{-} & \multirow{2}{*}{-}\\
                        &                                & $\rmS_3$: $2ML$  &                   &                   \\
\hline
\end{tabular}
\end{table} 

Fig. \ref{fig:CNN_Diag}(a) shows the architectures of the DE-CNN and RE-CNN, which combine the benefits of the convolutional layer (CL) and the \textcolor{black}{feedforward} layer (FFL).
As for CL, the neurons are locally connected, and a group of connections among them is assigned the same weight.
Thus, by introducing CL into a DL network, the network parameters are effectively decreased, and the convergence is accelerated compared with a fully connected network \cite{ref:CNN-merit}.
\textcolor{black}{The} FFL is further combined with the CL \textcolor{black}{to enhance} the estimation \textcolor{black}{accuracy} and \textcolor{black}{generalize the} ability of the CNNs.
In the developed DL framework, both DE-CNN and RE-CNN are composed of the input, hidden, and output layers.
Considering the aforementioned benefits, the hidden layers in the DE-CNN consist of one CL and one FFL with a flatten layer between them.
Since the propagation environment of the reflected \textcolor{black}{SAC} channels is more complicated, the hidden layers in the RE-CNN increase to two CLs and two FFLs to promote its feature extraction ability.
All the CLs utilize the {\textit {tanh}} activation function, while the FFLs adopt the {\textit {linear}} one.
The \textcolor{black}{\textit{Adam}} optimizer with a learning rate of $2\times 10^{-4}$ is employed to update the CNN parameters. 
The minibatch transitions of size $200$ are randomly selected to reduce the correlation of the training samples. 
Furthermore, a stopping criterion is adopted in the training process, \textcolor{black}{i.e.,} the training stops if the validation accuracy does not improve in $N_s=5$ consecutive epochs or the training epochs reach the maximum number of epochs $N_e=200$.
The \textcolor{black}{remaining} hyperparameters of the DE-CNN and RE-CNN are summarized in Table \ref{table:CNN}.

\newlength\myindent 
\setlength\myindent{2em}
\newcommand\bindent{
    \begingroup
    \setlength{\itemindent}{\myindent}
    \addtolength{\algorithmicindent}{\myindent} }
\newcommand\eindent{\endgroup}

\begin{algorithm}[!t]
\caption{CNN-based Channel Estimation Algorithm.}
{{\bf Initialize:} $t^{\rmS_\ell\rmI_k}=0$, $\ell\in\calN_1^3$, $k\in\calN_1^2$; } \\
{\bf Offline training:}

\begin{algorithmic}[1]  
  \FOR {$\rmS_\ell$, $\ell\in\calN_1^3$,}

  \FOR {$\rmI_k$, $k\in\calN_1^2$,}
    \STATE \textcolor{black}{{\bf Generate} training dataset $(\calT^{\rmS_\ell\rmI_k},\calO^{\rmS_\ell})$ according to Section \ref{sec:ChEst_IO} and \ref{sec:ChEst_training_set};}
    \STATE \textcolor{black}{{\bf Pre-process} $(\calT^{\rmS_\ell\rmI_k},\calO^{\rmS_\ell})$ by performing input standardization and output scaling;}
    \STATE {\bf Input} the pre-processed samples $(\bar\bfT^{\rmS_\ell\rmI_k}_{(v,u)},\bar\bfO^{\rmS_\ell}_{(v)})$ to the designed CNN $f^{\rmS_\ell\rmI_k}(\cdot \,; \Theta^{\rmS_\ell\rmI_k})$;




    \WHILE {$t^{\rmS_\ell\rmI_k}\leq N_e$ or the validation accuracy improves in $N_s$ consecutive epochs}
      \STATE {\bf Update} $\Theta^{\rmS_\ell\rmI_k}$ by minimizing ${\cal L}^{\rmS_\ell\rmI_k}$ in \eqref{eq:loss} using Adam optimizer;
      \STATE {\bf Set} $t^{\rmS_\ell\rmI_k} \leftarrow t^{\rmS_\ell\rmI_k}+1$;
    \ENDWHILE
    \STATE {\bf Output} the trained CNN $f^{\rmS_\ell\rmI_k}(\cdot \,; \hat\Theta^{\rmS_\ell\rmI_k})$.

  \ENDFOR

  \ENDFOR
\end{algorithmic}

{\bf Online estimation:}
\begin{algorithmic}[1]  

  \STATE {\bf Input:} Received signals $\bfy_c^{\rmS_\ell}$ that used to generate the testing dataset $\dot\calT^{\rmS_\ell\rmI_k}$, $\ell\in\calN_1^3$, $k\in\calN_1^2$;
  \begin{ALC@g}
    \STATE \textcolor{black}{{\bf Estimate} $\{\bfb, \bff\}$ by using the trained DE-CNN $f^{\rmS_1\rmI_k}(\cdot \,; \hat\Theta^{\rmS_1\rmI_k})$ with the standardized sample $\tilde{\bfT}^{\rmS_1\rmI_k}$;}
    \STATE \textcolor{black}{{\bf Estimate} $\bfG_\rmu$ by using the trained RE-CNN $f^{\rmS_2\rmI_k}(\cdot \,; \hat\Theta^{\rmS_2\rmI_k})$ with the standardized sample $\tilde{\bfT}^{\rmS_2\rmI_k}$;}
    \STATE \textcolor{black}{{\bf Estimate} $\bfG_\rmt$ by using the trained RE-CNN $f^{\rmS_3\rmI_k}(\cdot \,; \hat\Theta^{\rmS_3\rmI_k})$ with the standardized sample $\tilde{\bfT}^{\rmS_3\rmI_k}$;}
  \end{ALC@g}
  \STATE {\bf Output:} Estimated channels $\{\hat\bfb, \hat\bff, \hat\bfG_\rmu, \hat\bfG_\rmt\}$.

\end{algorithmic}
\label{algo:DL-CE}
\end{algorithm}

\subsubsection{Online Estimation}
\textcolor{black}{The online estimation phase for the three stages is illustrated in Fig. \ref{fig:CNN_Diag}(b), comprised of data pre-processing, channel estimation with the trained DL network, and data post-processing.
First, the testing dataset of the $k$-th type in the $\ell$-th stage, $\dot{\calT}^{\rmS_\ell\rmI_k}$, is pre-processed to derive the standardized samples as $\tilde{\bfT}^{\rmS_\ell\rmI_k}$.
Then, the output of the trained CNN, $\tilde{\bfO}^{\rmS_\ell}$, is obtained by
\begin{align}
\tilde{\bfO}^{\rmS_\ell} = f^{\rmS_\ell\rmI_k}(\tilde{\bfT}^{\rmS_\ell\rmI_k};\hat\Theta^{\rmS_\ell\rmI_k}),
\end{align}
where $\hat\Theta^{\rmS_\ell\rmI_k}$ denotes the hyperparameters of the trained CNN.
Further, by scaling $\tilde{\bfO}^{\rmS_\ell}$ with \textcolor{black}{a} factor \textcolor{black}{of} $\delta^{-1}$ and performing straightforward {mathematical operations (i.e., combine the real and imaginary parts in $\tilde{\bfO}^{\rmS_\ell}$ to the complex numbers)}, the channels of the direct {SAC}, reflected communication, and reflected sensing \textcolor{black}{channels} are successively estimated as $\{\hat\bfb,\hat\bff\}$, $\hat\bfG_\rmu$, and $\hat\bfG_\rmt$.}
The proposed CNN-based \textcolor{black}{SAC} channels estimation algorithm is summarized in Algorithm \ref{algo:DL-CE}, where $t^{\rmS_\ell\rmI_k}$ represents the index of the training epoch that corresponds to the $k$-th type of input-output pair in \textcolor{black}{$\rmS_\ell$}.


\section{Complexity Analysis}\label{sec:complexity}
This section discusses the computational complexity of the proposed approach \textcolor{black}{for} inputs generation and CNN-based online estimation.
The number of real additions and multiplications are \textcolor{black}{considered} to assess \textcolor{black}{the} complexity.

\subsection{Complexity of Inputs Generation}
Since the generation of the first type of inputs \textcolor{black}{(i.e., $\bfT^{\rmS_1\rmI_1}$, $\bfT^{\rmS_2\rmI_1}$, and $\bfT^{\rmS_3\rmI_1}$)} does not require signal pre-processing, \textcolor{black}{there is no additional} computational complexity.

For the generation of the second type of input in \textcolor{black}{$\rmS_1$} (i.e., $\bfT^{\rmS_1\rmI_2}$), the computational complexity comes from \eqref{eq:bar_b} and \eqref{eq:bar_f}.
Thus, the number of real additions and multiplications, $\calC_\calA^{\rmS_1\rmI_2}$ and $\calC_\calM^{\rmS_1\rmI_2}$, required to generate the input $\bfT^{\rmS_1\rmI_2}$ are given {respectively} by
\begin{align}
\calC_\calA^{\rmS_1\rmI_2} = \frac{10}{3} C^{\rmS_1}M(2M^2-1) - 2M,
\end{align}
and
\begin{align}
\calC_\calM^{\rmS_1\rmI_2} = \frac{2}{3} C^{\rmS_1}M(52M^2+39M-1) + 4M +2.
\end{align}

Regarding the generation of the second type of input in \textcolor{black}{$\rmS_2$} (i.e., $\bfT^{\rmS_2\rmI_2}$), the complexity is introduced by \eqref{eq:ytilde_cS2}, \eqref{eq:ybar_cS2}, and \eqref{eq:bar_Gu}.
The cost of \eqref{eq:ytilde_cS2} is ${\tilde \calC}_\calA^{\rmS_2\rmI_2}= 2M^2$ real additions and ${\tilde \calC}_\calM^{\rmS_2\rmI_2}=4M^2$ real multiplications in each sub-frame. 
Particularly, for the inverse calculation of a complex $q\times q$ matrix, the required number of real additions and multiplications are ${\calC}_\calA^{{\rm inv},q} = \frac 2 3q(3q^2+3q-1)$ and ${\calC}_\calM^{{\rm inv},q} = \frac 1 3 q(4q^2 + 15q - 1)$, respectively \cite{ref:matrix-inverse}.
Then, the cost of \eqref{eq:ybar_cS2} is ${\bar\calC}_\calA^{\rmS_2\rmI_2}=2M(2M^2-M-1) + {\calC}_\calA^{{\rm inv},M}$ real additions and ${\bar \calC}_\calM^{\rmS_2\rmI_2}=4M^2(2M+1) + {\calC}_\calM^{{\rm inv},M}$ real multiplications in each sub-frame.
The LS estimator in \eqref{eq:bar_Gu} costs ${\bar\calC}_\calA^{\rmu}=(4L^2-2L)(C^{\rmS_2}-C^{\rmS_1})-2L^2 +{\calC}_\calA^{{\rm inv},L}$ and ${\bar\calC}_\calM^{\rmu} = 8L^2(C^{\rmS_2}-C^{\rmS_1})+{\calC}_\calM^{{\rm inv},L}$ real additions and multiplications, respectively.
Therefore, the number of real additions and multiplications, $\calC_\calA^{\rmS_2\rmI_2}$ and $\calC_\calM^{\rmS_2\rmI_2}$, required to generate the input $\bfT^{\rmS_2\rmI_2}$ can be \textcolor{black}{expressed respectively} as 
\begin{align}
\calC_\calA^{\rmS_2\rmI_2} & =  (C^{\rmS_2}-C^{\rmS_1})({\tilde \calC}_\calA^{\rmS_2\rmI_2} + {\bar \calC}_\calA^{\rmS_2\rmI_2}) + {\bar\calC}_\calA^{\rmu} \notag \\
& = \frac 2 3 (C^{\rmS_2}-C^{\rmS_1})(9M^3+3M^2-4M +6L^2-3L) + \frac 2 3 L(3L^2-1),
\end{align}
and
\begin{align}
\calC_\calM^{\rmS_2\rmI_2} & = (C^{\rmS_2}-C^{\rmS_1})({\tilde \calC}_\calM^{\rmS_2\rmI_2} + {\bar \calC}_\calM^{\rmS_2\rmI_2}) + {\bar\calC}_\calM^{\rmu} \notag \\
& = \frac 1 3 (C^{\rmS_2}-C^{\rmS_1})(22M^3+39M^2-M  +24L^2) + \frac 1 3 (4L^2+15L+1).
\end{align}

The second type of input in \textcolor{black}{$\rmS_3$} (i.e., $\bfT^{\rmS_3\rmI_2}$) is generated according to \eqref{eq:ytilde_cS3}, \eqref{eq:ybar_cS3}, and \eqref{eq:bar_Gt}.
In each sub-frame, the cost of \eqref{eq:ytilde_cS3} is ${\tilde \calC}_\calA^{\rmS_3\rmI_2}=2M(2M+L)$ real additions and ${\tilde \calC}_\calM^{\rmS_3\rmI_2}=4M(2M+L)$ real multiplications.
With straightforward derivation, the complexity of \eqref{eq:ybar_cS3} and \eqref{eq:bar_Gt} is respectively equal to \eqref{eq:ybar_cS2} and \eqref{eq:bar_Gu}.
Hence, the number of real additions and multiplications, $\calC_\calA^{\rmS_3\rmI_2}$ and $\calC_\calM^{\rmS_3\rmI_2}$, required to generate $\bfT^{\rmS_3\rmI_2}$ are given \textcolor{black}{respectively} by 
\begin{align}
\calC_\calA^{\rmS_3\rmI_2}
& = \frac 2 3 (C^{\rmS_3}-C^{\rmS_2})(9M^3+6M^2-4M+3ML  +6L^2-3L) + \frac 2 3 (3L^2-1),
\end{align}
and
\begin{align}
\calC_\calM^{\rmS_3\rmI_2}
& = \frac 1 3 (C^{\rmS_3}-C^{\rmS_2})(22M^3+51M^2-M+12ML  +24L^2) + \frac 1 3 L(4L^2+15L-1).
\end{align}

\subsection{Complexity of DE-CNN and RE-CNN}

\textcolor{black}{Referring to the proposed CNN-based estimation framework, the computational complexity of the proposed DE-CNN and RE-CNN comes from the {computational complexity} calculation of each layer.
To formulate the general expressions of the computational complexity, {consider} $\eta_i$ and $F_{n_i}$ {as} the number of neurons and filters in the $i$-th layer, {respectively}, while $F_z$ denotes the filter size of the CL.
Their exact values are presented in Table \ref{table:CNN}.
For computational simplicity, the cost of each activation function is considered as one real addition.}

\textcolor{black}{Regarding the proposed DE-CNN, the computational complexity is introduced by the \textcolor{black}{computational complexity} calculation of the second, third, and fourth layers.
Their required number of real additions is respectively expressed as $F_{n_2}\eta_{F_2}(F_z+1)$, $\eta_3(F_{n_2}\eta_{F_2}+1)$, and $\eta_4(\eta_3+1)$, whereas that of the real multiplications is respectively given by $F_{n_2}\eta_{F_2}F_z$, $F_{n_2}\eta_{F_2}\eta_3$, and $\eta_3\eta_4$.
Here, $\eta_{F_2} = \lfloor {\frac{\eta_1-F_z} {F_s} + 1}\rfloor$ represents the output size of one filter in the second CL, and $F_s=1$ is the stride.}
Therefore, the required number of real additions and multiplications, $\calC_\calA^{\rm DE}$ and $\calC_\calM^{\rm DE}$, {for} the DE-CNN are given {respectively} by
\begin{align}
\calC_\calA^{\rm DE} = F_{n_2}\eta_{F_2}(F_z+\eta_3+1) + \eta_4(\eta_3+1) + \eta_3,
\end{align}
and
\begin{align}
\calC_\calM^{\rm DE} = F_{n_2}\eta_{F_2}(F_z+\eta_3) + \eta_3\eta_4.
\end{align}

\textcolor{black}{Similarly, the computational complexity of the proposed RE-CNN comes from the \textcolor{black}{computational complexity} calculation of the second to the sixth layers.
For both the second and third CLs (i.e., $i\in\calN_2^3$), the required number of real additions and multiplications are  $F_{n_i}\eta_{F_i}(F_z+1)$ and $F_{n_i}\eta_{F_i}F_z$, respectively.
Define $\eta_{F_3} = \lfloor {\frac{F_{n_2}\eta_{F_2}-F_z} {F_s} + 1} \rfloor$ as the output size of one filter in the third CL.
The cost of the \textcolor{black}{computational complexity} calculation for the fourth layer is $\eta_4(F_{n_3}\eta_{F_3}+1)$ real additions and $F_{n_3}\eta_{F_3}\eta_4$ real multiplications.
For both the fifth and sixth layers (i.e., $i\in\calN_5^6$), the \textcolor{black}{computational complexity} calculation costs $\eta_i(\eta_{i-1}+1)$ real additions and $\eta_i\eta_{i-1}$ real multiplications.
Hence, the required number of real additions and multiplications, $\calC_\calA^{\rm RE}$ and $\calC_\calM^{\rm RE}$, {for} the RE-CNN can be written {respectively} as
\begin{align}
& \calC_\calA^{\rm RE} = \sum_{i\in\calN_2^3}  F_{n_i} \eta_{F_i} (F_z + 1) + \sum_{i\in\calN_5^6} \eta_{i} (\eta_{i-1} + 1)  + \eta_4 (F_{n_3}\eta_{F_3} + 1),
\end{align}
and
\begin{align}
\calC_\calM^{\rm RE} = \sum_{i\in\calN_2^3} F_{n_i} \eta_{F_i} F_z + \sum_{i\in\calN_5^6} \eta_i\eta_{i-1} + F_{n_3} \eta_{F_3} \eta_4.
\end{align}}

\section{Simulation Results}\label{sec:simulation}

This section quantitatively evaluates the performance of the proposed three-stage channel estimation approach for the IRS-assisted ISAC system, considering the LS estimator as the baseline scheme.
The simulation setup is firstly provided.
Then, the channel estimation performance of the proposed approach is unveiled under various SNR conditions and system parameters.
Ultimately, the computational complexity of the proposed approach in terms of different system parameters is investigated.

\subsection{Simulation Setup}
\begin{figure}
\centering
\includegraphics[width=2.8in]{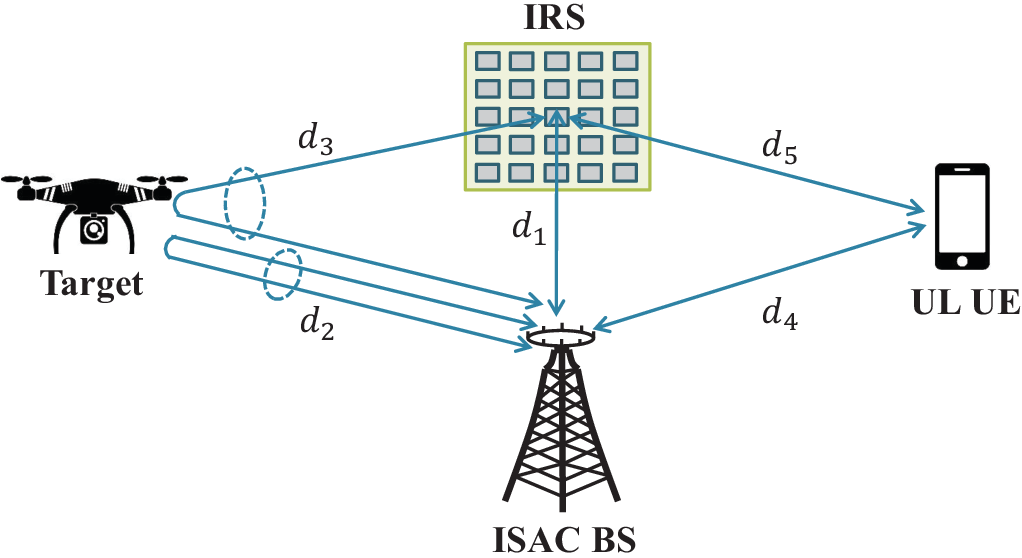}
\caption{Simulation setup.}
\label{fig:System_distance}
\end{figure}

\textcolor{black}{In simulations, $M = 4$, $L=30$, $C^{\rmS_1}=1$, $C^{\rmS_2}=L+1$, and $C^{\rmS_3}=2L+1$ unless further specified.}
As in \cite{ref:ChModel-b}, the sensing channel of the BS-target-IRS link is modeled as 
\begin{align}
\bfA = \alpha_1\bfa(\theta_{\rm BT})\bfa(\theta_{\rm TI})^{\rmH},
\label{eq:radar_model}
\end{align}
where $\alpha_1$ denotes the complex-valued reflection coefficient of the BS-target-IRS link, and its phase-shift is uniformly distributed from $[0,2\pi)$ \cite{ref:ChModel-b,ref:channel-b-fixed-amplitude}.
For the $\bar M$ array elements, define the steering vector $\bfa(\bar\theta)$ associated with the azimuth angle $\bar\theta$ as
\begin{align}
\bfa(\bar\theta) = [1, e^{\jmath\frac{2\pi {\bar d}}{\lambda}\sin(\bar\theta)}, \ldots, e^{\jmath\frac{2\pi {\bar d}}{\lambda}({\bar M}-1)\sin(\bar\theta)}]^\rmT,
\label{eq:steering_vec}
\end{align}
where $\bar d$ and $\lambda$ are the inter-element spacing and signal wavelength, respectively.
Then, $\bfa(\theta_{\rm BT})$ and $\bfa(\theta_{\rm TI})$ in \eqref{eq:radar_model} can be obtained according to \eqref{eq:steering_vec}.
Let $\bar\theta=\theta_{\rm BT}$, $\bar d = d_\rmB$, and $\bar M=M$ in $\bfa(\theta_{\rm BT})$, while $\bar\theta=\theta_{\rm TI}$, $\bar d = d_\rmI$, and $\bar M=L$ in $\bfa(\theta_{\rm TI})$.
Here, $\theta_{\rm BT}$ and $\theta_{\rm TI}$ represent the azimuth angles of the target and IRS, respectively associated with the BS-target and target-IRS links.
\textcolor{black}{Without loss of generality}, the antenna spacing of \textcolor{black}{the} ISAC BS and distance between two adjacent IRS elements, $d_\rmB$ and $d_\rmI$, satisfy $d_\rmB=d_\rmI=\frac{\rho}{2}$.
Similar to \eqref{eq:radar_model}, the BS-target-BS channel is modeled as $\bfb=\alpha_2\bfa(\theta_{\rm BT})$ with $\alpha_2$ as its reflection coefficient, which follows the same distribution as $\alpha_1$ \textcolor{black}{with} $|\alpha_1|=|\alpha_2|=1$.
The rest of the channels are modeled as Rician fading \cite{ref:DL-IRS-ChE-TWC}.
As such, the channel of the IRS-BS link is given by
\begin{align}
\bfg = \sqrt{\frac {K_\rmIB}{K_\rmIB+1}} \bfg_{\rmLoS}
+ \sqrt{\frac {1}{K_\rmIB+1}} \bfg_{\rmNLoS},
\label{eq:g}
\end{align}
where $K_\rmIB=10$ is the Rician factor of the IRS-BS channel.
Here, $\bfg_{\rmLoS}$ and $\bfg_{\rmNLoS}$ respectively denote the line-of-sight (LoS) component and non-LoS (NLoS) component of $\bfg$.
By utilizing \eqref{eq:steering_vec}, define $\bfg_{\rmLoS}=\bfa(\theta_{\rm IB})^{\rmH}$, where $\theta_{\rm IB}$ is the azimuth angle from the IRS to the BS.
The communication channels \textcolor{black}{(i.e., $\bff$ and $\bfH$)} can be modeled similarly as in \eqref{eq:g}, and their Rician factors are set to $0$ \textcolor{black}{(i.e., Rayleigh fading)}.
Furthermore, a distance-dependent path loss model is adopted to characterize the path loss factor of each channel.
As shown in Fig. \ref{fig:System_distance}, the path loss at distance $d_j$, $j\in\calN_1^5$, is modeled as $\rho_j = \rho_0(\frac{d_j}{d_0})^{-\gamma_j}$, where $\rho_0=-30\,{\rm dBm}$ is the path loss at the reference distance $d_0=1\,\rm m$\textcolor{black}{,} and $\gamma_j$ denotes the path loss exponent \cite{ref:ChModel-refpower,ref:ChModel-power}. 
The distances are set to $d_1=2\,\rm m$, $d_2=d_3=140\,\rm m$, and $d_4=d_5=50\,\rm m$.
The azimuth angles are $\theta_{\rm BT}=\arccos(\frac{d_1}{d_5})$, $\theta_{\rm TI}=-\arccos(\frac{d_1}{d_5})$, and $\theta_{\rm IB}=\pi$.
The corresponding path loss exponents are $\gamma_1=2$, $\gamma_2=3.5$, $\gamma_3=2.3$, $\gamma_4=3$, and $\gamma_5=2.2$ \cite{ref:ChModel-refpower}.
Referring to \cite{ref:ChModel-power}, the transmit power of the ISAC BS and UL UE is set to $\mathcal P_\rmB=20\,\rm dBm$ and $\mathcal P_\rmU=15\,\rm dBm$, respectively.

\begin{figure}
\centering
\subfigure[ ]
{\begin{minipage}[b]{0.32\textwidth}
\includegraphics[width=1\textwidth]{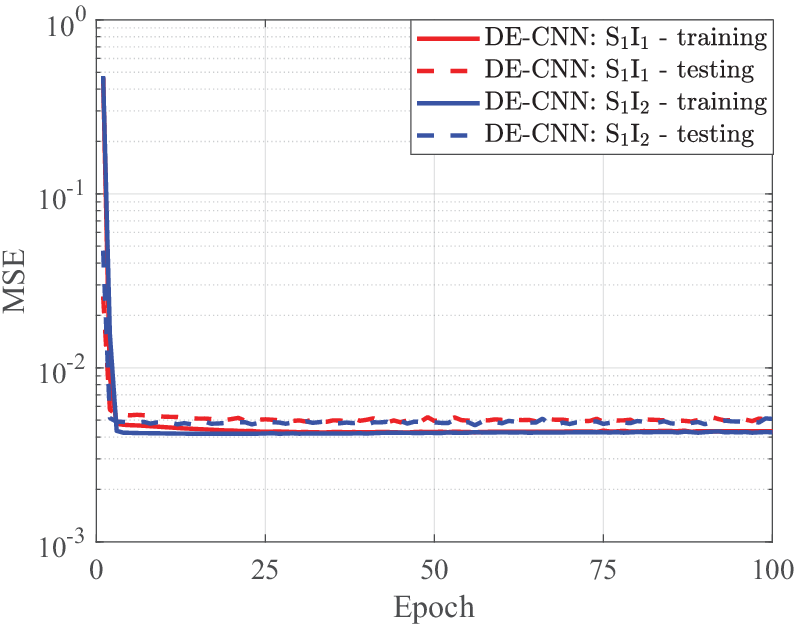}
\end{minipage}}
\subfigure[ ]
{\begin{minipage}[b]{0.32\textwidth}
\includegraphics[width=1\textwidth]{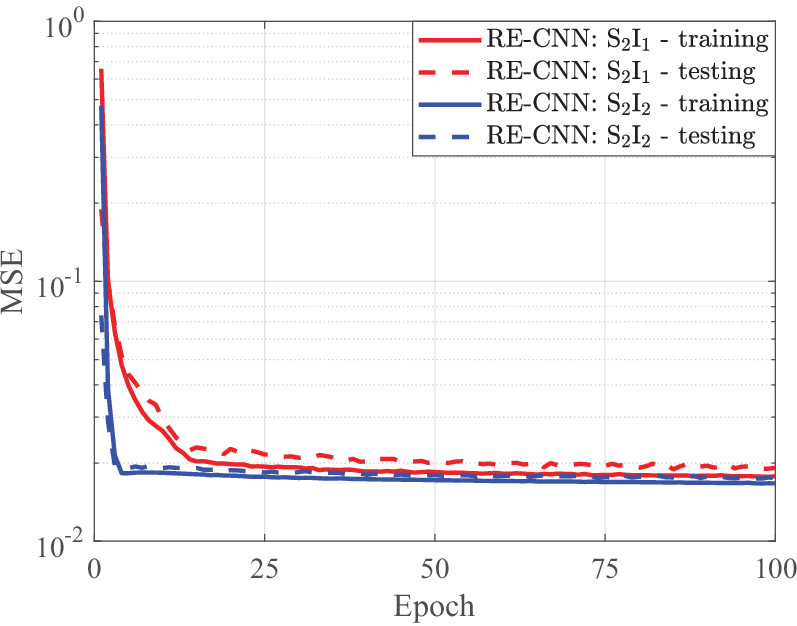}
\end{minipage}}
\subfigure[ ]
{\begin{minipage}[b]{0.32\textwidth}
\includegraphics[width=1\textwidth]{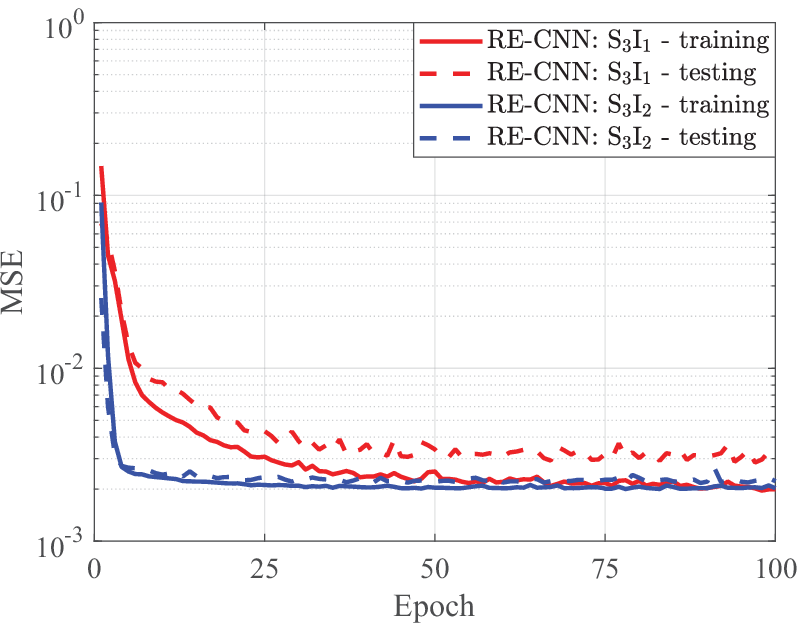}
\end{minipage}}
\caption{MSE of CNNs training and testing versus epoch for $M=4$, $L=15$, and $\text{SNR}=15\rm\,dB$: (a) DE-CNN for $\rmS_1$, (b) RE-CNN for $\rmS_2$, (c) RE-CNN for $\rmS_3$.}
\label{fig:CNN_epoch}
\end{figure}

The hyperparameters of the \textcolor{black}{proposed} DE-CNN and RE-CNN \textcolor{black}{are} summarized in Table \ref{table:CNN}.
\textcolor{black}{For} the offline training phase, the training dataset size is set to $T_{\rm off}=VU=10^4$ for each SNR condition with $V=10^3$, $U=10$, and $\text{SNR}_\text{ch}=30\,\rm dB$.
Adopt $90\%$ of the dataset for training and the remaining \textcolor{black}{dataset} for testing.
As seen from Fig. \ref{fig:CNN_epoch}, the convergence properties of the \textcolor{black}{proposed} CNNs (i.e., DE-CNN for $\rmS_1$ and RE-CNN for $\rmS_2$ and $\rmS_3$) are evaluated by using the data samples generated \textcolor{black}{at} $\text{SNR}=15\,\rm dB$.
It is shown that the DE-CNN and RE-CNN trained by the different types of input-output pairs \textcolor{black}{provide} fast convergence, while the MSE in terms of training and testing are extremely close.
\textcolor{black}{For} the online estimation phase, \textcolor{black}{different} $T_{\rm on}=10^3$ data samples are tested to verify the estimation performance for \textcolor{black}{various} SNR \textcolor{black}{conditions}.
The SNR used in the simulations is defined as $\text{SNR} = \frac{{\cal P}_{\mathrm r} }{\sigma^2}$, where ${\cal P}_{\mathrm r}$ and $\sigma^2$ respectively represent the received signal power and noise power at the ISAC BS.
\textcolor{black}{For the three stages (i.e., $\rmS_1$, $\rmS_2$, and $\rmS_3$)}, ${\cal P}_{\mathrm r}$ equals ${\cal P}_\rmB\rho_2+{\cal P}_\rmU\rho_4$, ${\cal P}_\rmU(\rho_4+\rho_1\rho_5)$, and ${\cal P}_\rmB(\rho_2+\rho_1\rho_3)+{\cal P}_\rmU(\rho_4+\rho_1\rho_5)$, respectively.
To assess the estimation performance, the normalized MSE (NMSE) is \textcolor{black}{utilized} as \textcolor{black}{a} performance metric\textcolor{black}{, and} is given by
\begin{align}
\text{NMSE}=\mathbb{E}\bigg\{ \frac{\| \text{Estimated} - \text{True} \|_F^2}{\| \text{True} \|_F^2} \bigg\}.
\label{eq:NMSE}
\end{align}

\subsection{NMSE versus SNR}

\begin{figure}
\centering
\subfigure[ ]
{\begin{minipage}[b]{0.32\textwidth}
\includegraphics[width=1\textwidth]{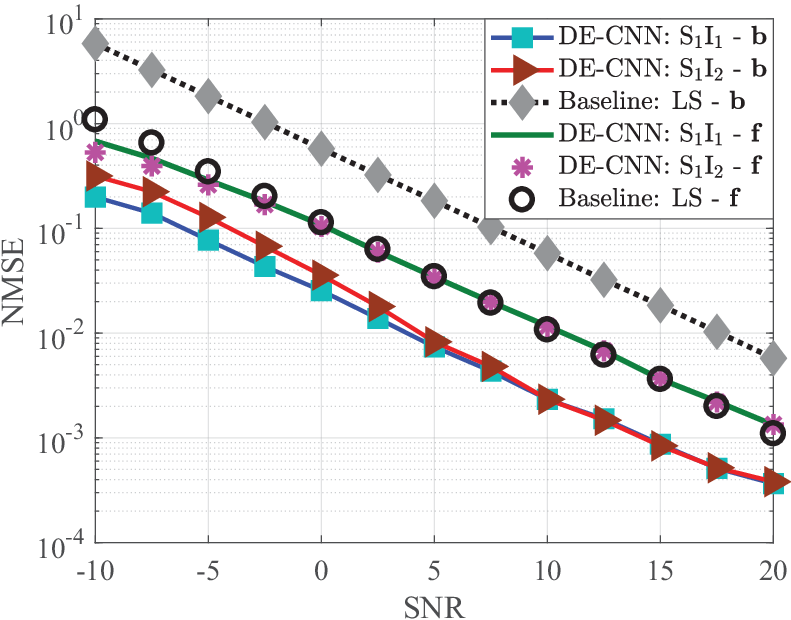}
\end{minipage}}
\subfigure[ ]
{\begin{minipage}[b]{0.32\textwidth}
\includegraphics[width=1\textwidth]{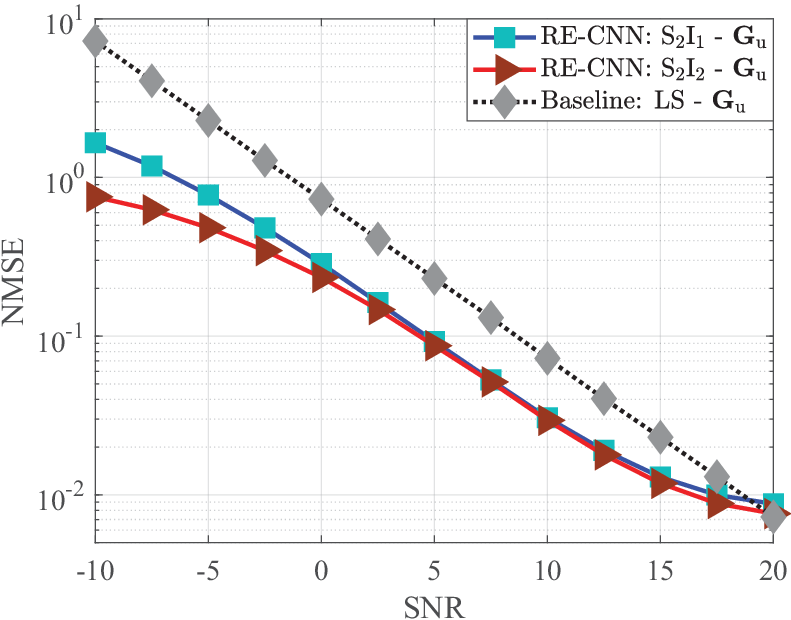}
\end{minipage}}
\subfigure[ ]
{\begin{minipage}[b]{0.32\textwidth}
\includegraphics[width=1\textwidth]{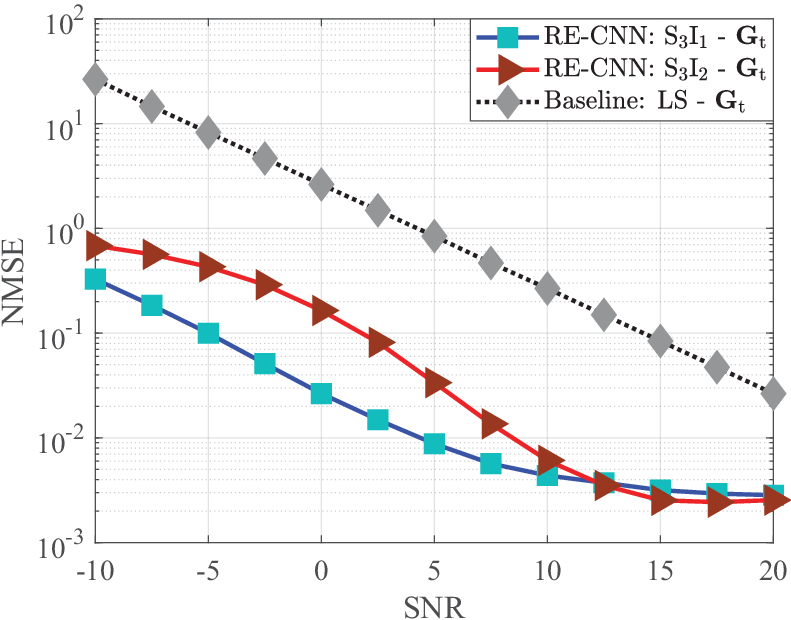}
\end{minipage}}
\caption{NMSE of \textcolor{black}{SAC} channels estimation versus SNR for $M=4$ and $L=30$: (a) Direct \textcolor{black}{SAC} channels, (b) Reflected communication channel, (c) Reflected sensing channel.}
\label{fig:NMSE_SNR}
\end{figure}

Fig. \ref{fig:NMSE_SNR} illustrates the impact of SNR on the NMSE performance for the proposed \textcolor{black}{DL} approach and \textcolor{black}{LS} baseline scheme.
\textcolor{black}{Consider} $\text{SNR} = [10,20]\,\rm{dB}$ with a step size of $5\,\rm{dB}$ in the offline training phase, while $\text{SNR} = [-10,20]\,\rm{dB}$ with a step size of $2.5\,\rm{dB}$ for online estimation.

The NMSE performance of the direct \textcolor{black}{SAC} channels estimation is depicted in Fig. \ref{fig:NMSE_SNR}(a).
As can be observed, for both $\bfb$ and $\bff$, the proposed \textcolor{black}{DL} approach \textcolor{black}{provides a} comparable estimation performance \textcolor{black}{to the LS baseline scheme} when trained by the different types of input-output pairs (i.e., $\big(\bfT^{\rmS_1\rmI_1},\bfO^{\rmS_1}\big)$ and $\big(\bfT^{\rmS_1\rmI_2}, \bfO^{\rmS_1}\big)$), while it is superior to the \textcolor{black}{LS} baseline scheme.
One can also note that compared to the \textcolor{black}{LS} baseline scheme, the proposed \textcolor{black}{DL} approach attains $12.5\,\rm dB$ SNR improvement at $\text{NMSE}=10^{-1}$ for \textcolor{black}{estimating} $\bfb$.
\textcolor{black}{The} performance improvement is more noticeable than that for \textcolor{black}{estimating} $\bff$.
The reason is that the sensing channel model of $\bfb$ is simpler than the communication one for $\bff$, and thus, the mapping of the input-output \textcolor{black}{pairs} for $\bfb$ is \textcolor{black}{easier} to be learned by the proposed DE-CNN.

Based on the estimated $\hat\bff$ \textcolor{black}{at} $\rmS_1$, Fig. \ref{fig:NMSE_SNR}(b) \textcolor{black}{shows} the estimation performance \textcolor{black}{of} the reflected communication channel\textcolor{black}{,} $\bfG_\rmu$.
It is shown that the NMSE performance of the proposed \textcolor{black}{DL} approach is comparable \textcolor{black}{to the LS baseline scheme} when trained by the different types of input-output pairs (i.e., $\big(\bfT^{\rmS_2\rmI_1}, \bfO^{\rmS_2}\big)$ and $\big(\bfT^{\rmS_2\rmI_2}, \bfO^{\rmS_2}\big)$), whereas it significantly outperforms the \textcolor{black}{LS} baseline scheme, especially in the low SNR region.
For instance, the proposed \textcolor{black}{DL} approach achieves around $4\,\rm dB$ SNR improvement at $\text{NMSE}=10^{-1}$ compared to the \textcolor{black}{LS} baseline scheme.

Fig. \ref{fig:NMSE_SNR}(c) evaluates the estimation performance \textcolor{black}{of} the reflected sensing channel $\bfG_\rmt$, based on the estimated $\{\hat\bfb,\hat\bff\}$ and $\hat\bfG_\rmu$ from the previous stages.
By adopting the second type of input-output pair $\big(\bfT^{\rmS_3\rmI_2}, \bfO^{\rmS_3}\big)$, the proposed \textcolor{black}{DL} approach achieves up to $12.5\,\rm dB$ SNR improvement at $\text{NMSE}=10^{-1}$ compared to the \textcolor{black}{LS} baseline scheme.
Moreover, $7\,\rm dB$ SNR improvement is obtained by the proposed \textcolor{black}{DL} approach when using the first type of input-output pair $\big(\bfT^{\rmS_3\rmI_1}, \bfO^{\rmS_3}\big)$.
This \textcolor{black}{lies} in that the proposed RE-CNN can correct partial estimation errors introduced \textcolor{black}{by} $\rmS_1$ and $\rmS_2$ if trained by $\big(\bfT^{\rmS_3\rmI_1}, \bfO^{\rmS_3}\big)$. 
Note that the generation of $\bfT^{\rmS_3\rmI_1}$ builds on the original received signals without \textcolor{black}{a need of} signal pre-processing (i.e., $\bfy_c^{\rmS_3}$), as formulated in \eqref{eq:I_S3I1}.
On the contrary, when trained by $\big(\bfT^{\rmS_3\rmI_2}, \bfO^{\rmS_3}\big)$, the RE-CNN \textcolor{black}{cannot correct} the previous estimation errors since $\bfT^{\rmS_3 \rmI_2}$ in \eqref{eq:I_S3I4} relies on the \textcolor{black}{LS estimation results (i.e., ${\bar \bfG}_\rmt$ in \eqref{eq:bar_Gt})}.
These rough \textcolor{black}{estimations} have been corrupted by the previous estimation errors during signal pre-processing, thus, affecting the estimation accuracy.
\textcolor{black}{By taking a deep look at Figs. \ref{fig:NMSE_SNR}(b) and \ref{fig:NMSE_SNR}(c), one can note that compared to the LS baseline scheme, the performance improvement of the proposed DL approach is more noticeable for estimating $\bfG_\rmt$ than that for estimating $\bfG_\rmu$.
This is due to the fact that the sensing channel model of $\bfG_\rmt$ is simpler than the communication one \textcolor{black}{of} $\bfG_\rmu$.
The mapping of the input-output pairs \textcolor{black}{of} $\bfG_\rmt$ is more straightforward to be trained by the proposed RE-CNN, and the estimation accuracy \textcolor{black}{of} $\bfG_\rmt$ is further enhanced.}
In addition to the outstanding NMSE performance, Fig. \ref{fig:NMSE_SNR} also unveils that the proposed \textcolor{black}{DL} approach exhibits \textcolor{black}{an} excellent generalization capacity since the CNNs are trained for several specific SNR conditions that can be extended to a wide range of SNR regions.

\subsection{NMSE versus Channel Dimension}
\begin{figure}
\centering
\subfigure[ ]
{\begin{minipage}[b]{0.34\textwidth}
\includegraphics[width=1\textwidth]{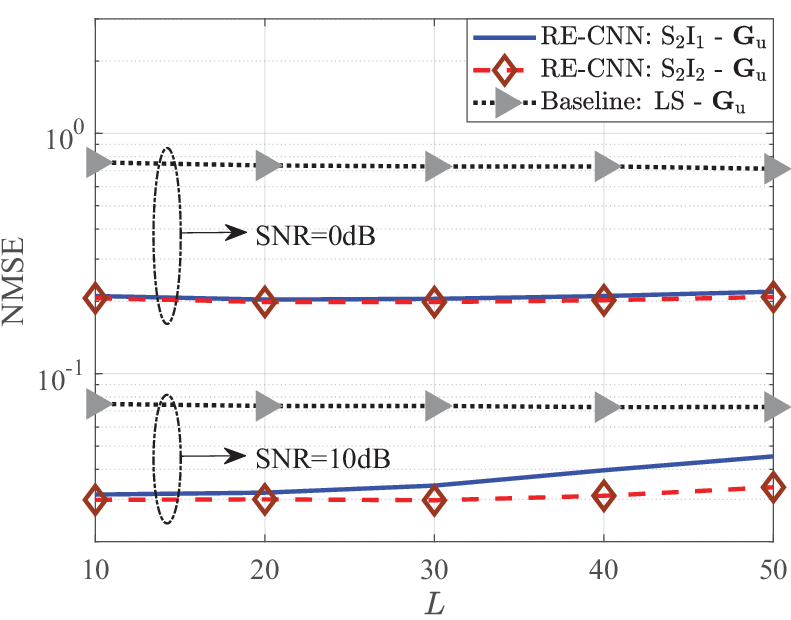}
\end{minipage}}
\subfigure[ ]
{\begin{minipage}[b]{0.34\textwidth}
\includegraphics[width=1\textwidth]{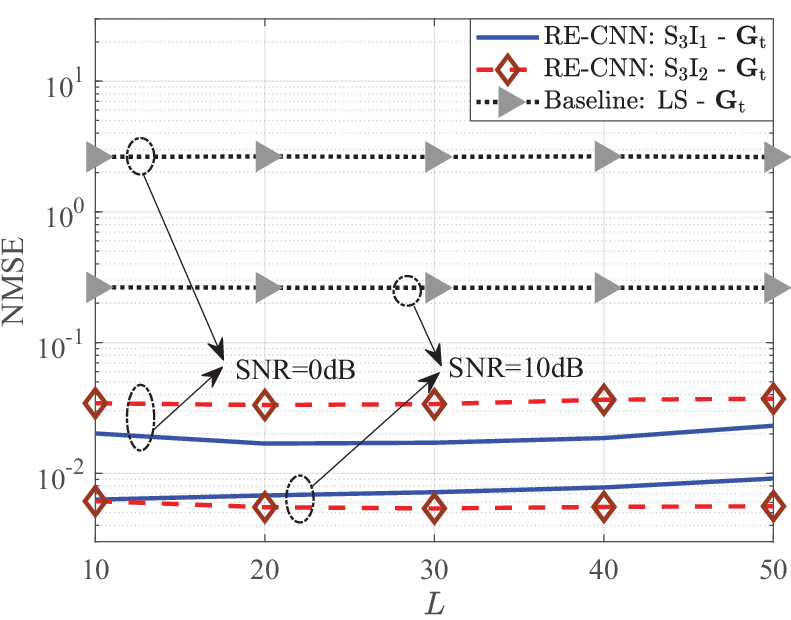}
\end{minipage}}

\caption{NMSE of \textcolor{black}{SAC} channels estimation versus $L$ at $M=4$ for $\text{SNR}=0\,\rm dB$ and $10\,\rm dB$: (a) Reflected communication channel, (b) Reflected sensing channel.}
\label{fig:NMSE_L}
\end{figure}

Note that $L$ directly affects the channel dimension of $\bfG_\rmu$ and $\bfG_\rmt$.
Therefore, Fig. \ref{fig:NMSE_L} illustrates the impact of increasing $L$ on the estimation performance.
The SNR in the offline training and online estimation phases is fixed to $0\,\rm dB$ and $10\,\rm dB$, \textcolor{black}{respectively}.

\begin{figure}
\centering
\subfigure[ ]
{\begin{minipage}[b]{0.32\textwidth}
\includegraphics[width=1\textwidth]{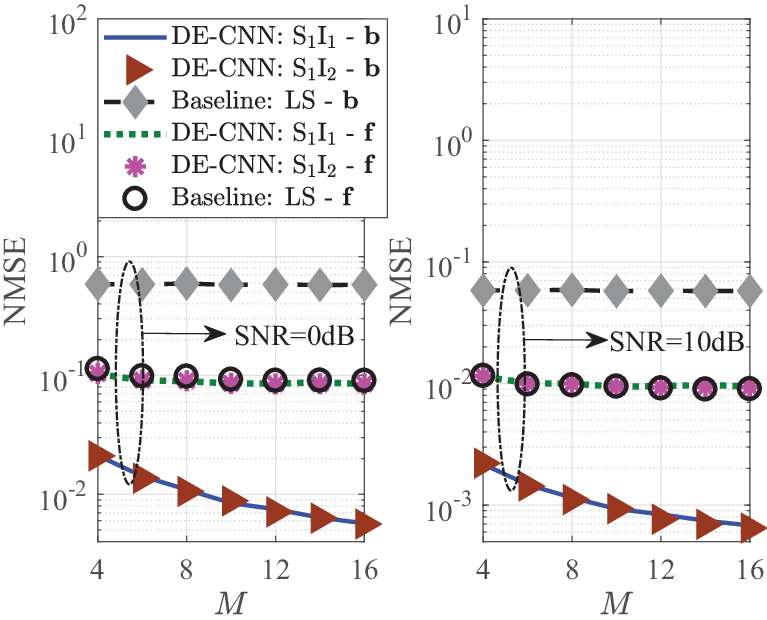}
\end{minipage}}
\subfigure[ ]
{\begin{minipage}[b]{0.32\textwidth}
\includegraphics[width=1\textwidth]{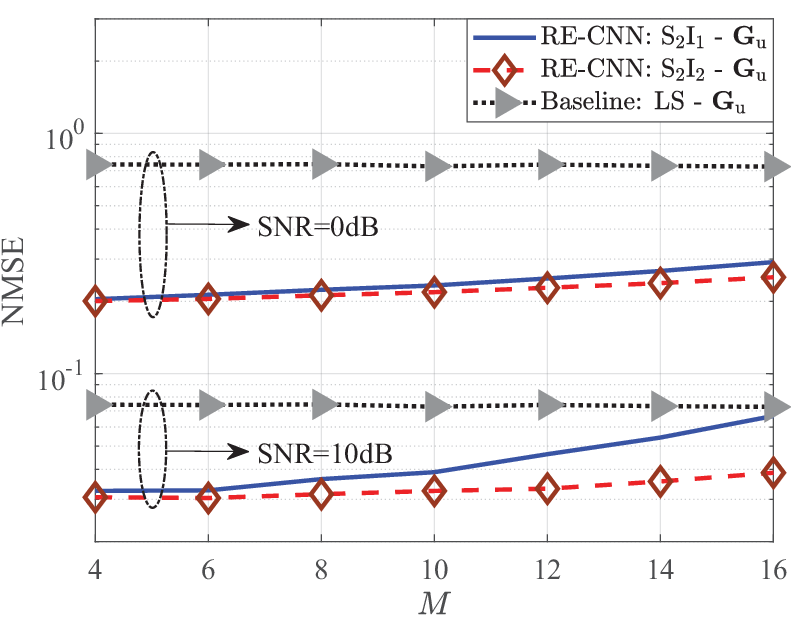}
\end{minipage}}
\subfigure[ ]
{\begin{minipage}[b]{0.32\textwidth}
\includegraphics[width=1\textwidth]{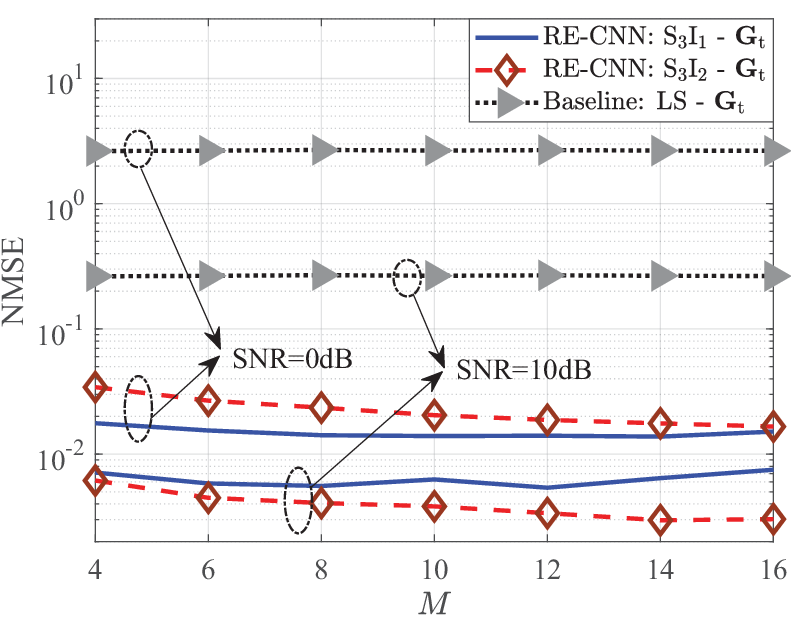}
\end{minipage}}
\caption{NMSE of \textcolor{black}{SAC} channels estimation versus $M$ at $L=15$ for $\text{SNR}=0\,\rm dB$ and $10 \,\rm dB$: (a) Direct \textcolor{black}{SAC} channels, (b) Reflected communication channel, (c) Reflected sensing channel.}
\label{fig:NMSE_M}
\end{figure}

Fig. \ref{fig:NMSE_L}(a) \textcolor{black}{depicts} the NMSE performance of the reflected communication channel $\bfG_\rmu$.
It is obvious that the proposed \textcolor{black}{DL} estimation approach attains significant performance enhancement compared to the \textcolor{black}{LS} baseline scheme for different $L$ values and SNR conditions.
\textcolor{black}{The proposed {DL} approach trained by the second type of input-output pair is robust to the change of $L$, {whereas that trained by the first one} slightly increases as $L$ increases.} 
This \textcolor{black}{lies} in the fact that the generation of the first input $\bfT^{\rmS_2 \rmI_1}$ \textcolor{black}{does not require a} signal pre-processing, and thus, the mapping of $\big(\bfT^{\rmS_2\rmI_1}, \bfO^{\rmS_2}\big)$ is more \textcolor{black}{challenging} to be learned accurately by the RE-CNN when the channel dimension becomes larger.

Fig. \ref{fig:NMSE_L}(b) illustrates the estimation performance of the reflected sensing channel $\bfG_\rmt$.
Apparently, the proposed \textcolor{black}{DL} approach outperforms the \textcolor{black}{LS} baseline scheme under different $L$ and SNR setups, and is robust to the change of $L$.
One can also note that the proposed {DL} approach {using} the first type of input-output pair {provides a} better {estimation accuracy} than that with the second one {at} $\text{SNR}=0\,\rm dB$, while the opposite \textcolor{black}{occurs} for $\text{SNR}=10\,\rm dB$. 
The former is attributed to the error correction ability of the {proposed} RE-CNN when trained by $\big(\bfT^{\rmS_3\rmI_1}, \bfO^{\rmS_3}\big)$, same as the findings in Fig. \ref{fig:NMSE_SNR}(c).
The latter lies in that the estimation errors from the previous stages are minor in the high SNR region and the mapping of $\big(\bfT^{\rmS_3 \rmI_2}, \bfO^{\rmS_3}\big)$ is \textcolor{black}{easy} to be learned by the \textcolor{black}{proposed} RE-CNN, thus, enhancing its NMSE performance.

To sum up, the above essential observations in Fig. \ref{fig:NMSE_L} unveil that the first type of input-output pairs (i.e., $\big(\bfT^{\rmS_2\rmI_1}, \bfO^{\rmS_2}\big)$ and $\big(\bfT^{\rmS_3\rmI_1}, \bfO^{\rmS_3}\big)$) are preferable in the estimation cases of \textcolor{black}{harsh} SNR conditions and \textcolor{black}{small} channel dimensions, whereas the second ones (i.e., $\big(\bfT^{\rmS_2\rmI_2}, \bfO^{\rmS_2}\big)$ and $\big(\bfT^{\rmS_3\rmI_2}, \bfO^{\rmS_3}\big)$) are recommended for the opposite \textcolor{black}{scenarios}.

Fig. \ref{fig:NMSE_M} studies the impact of varying $M$ on the estimation performance.
The SNR conditions are the same as \textcolor{black}{in} Fig. \ref{fig:NMSE_L} with $L=15$.
The NMSE performance of the direct \textcolor{black}{SAC} channels {(i.e., $\bfb$ and $\bff$)} is investigated in Fig. \ref{fig:NMSE_M}(a).
As can be observed, the proposed \textcolor{black}{DL} approach \textcolor{black}{performance} is superior to the \textcolor{black}{LS} baseline scheme for \textcolor{black}{estimating} $\bfb$, while \textcolor{black}{its performance is} comparable \textcolor{black}{with the LS baseline scheme} for \textcolor{black}{estimating} $\bff$.
Moreover, by adopting the proposed \textcolor{black}{DL} approach, the NMSE performance of \textcolor{black}{estimating} $\bfb$ decreases as $M$ increases.
This {lies} \textcolor{black}{in the fact} that the proposed DE-CNN can extract more distinguishable features of $\bfb$ to \textcolor{black}{improve} the estimation accuracy when $M$ becomes larger.

Figs. \ref{fig:NMSE_M}(b) and (c) respectively assess the estimation performance of the reflected \textcolor{black}{SAC} channels {(i.e., $\bfG_\rmu$ and $\bfG_\rmt$)}.
Obviously, for the estimation of $\bfG_\rmu$ and $\bfG_\rmt$, the proposed \textcolor{black}{DL} approach outperforms the \textcolor{black}{LS} baseline scheme under different $M$ values and SNR conditions.
Other findings from these two figures are similar as in Figs. \ref{fig:NMSE_L}(a) and (b).
Particularly, the NMSE of {estimating} $\bfG_\rmu$ achieved by the proposed \textcolor{black}{DL} approach is more robust to the change of $M$ when the RE-CNN is trained by the second type of input-output pair, as seen in Fig. \ref{fig:NMSE_M}(b).
For the estimation of $\bfG_\rmt$ in Fig. \ref{fig:NMSE_M}(c), the proposed \textcolor{black}{DL} approach trained by the first type of input-output pair (i.e., $\big(\bfT^{\rmS_3\rmI_1}, \bfO^{\rmS_3}\big)$) is superior to that with the second one (i.e., $\big(\bfT^{\rmS_3\rmI_2}, \bfO^{\rmS_3}\big)$) \textcolor{black}{at} $\text{SNR}=0\,\rm dB$, \textcolor{black}{and vice versa at} $\text{SNR}=10\,\rm dB$.

By taking an in-depth look at Figs. \ref{fig:NMSE_M}(c) and \ref{fig:NMSE_L}(b), one can also note that when the proposed \textcolor{black}{DL} approach adopts $\big(\bfT^{\rmS_3\rmI_2}, \bfO^{\rmS_3}\big)$, the NMSE of \textcolor{black}{estimating} $\bfG_\rmt$ slightly decreases as $M$ increases in Fig. \ref{fig:NMSE_M}(c), whereas it is stable for different $L$ values in Fig. \ref{fig:NMSE_L}(b).
This interesting finding {is} attributed to two aspects, as follows.
First, the increase of $L$ affects both BS-target-IRS and IRS-BS links {(i.e., $\bfA$ and $\bfg$)} of $\bfG_\rmt=\bfA\, \rmdiag\{{\bfg^\rmH}\}$, while the increase of $M$ only impacts $\bfA$.
This makes the channel estimation simpler when increasing $M$ \textcolor{black}{rather} than $L$.
Second, since the mapping of $\big(\bfT^{\rmS_3\rmI_2}, \bfO^{\rmS_3}\big)$ is \textcolor{black}{easier} than $\big(\bfT^{\rmS_3\rmI_1}, \bfO^{\rmS_3}\big)$, the proposed RE-CNN trained by $\big(\bfT^{\rmS_3\rmI_2}, \bfO^{\rmS_3}\big)$ {attains} \textcolor{black}{a} superior feature extraction ability \textcolor{black}{and enhances} the estimation accuracy of $\bfG_\rmu$ when $M$ becomes larger.

\subsection{Complexity Assessment}

\begin{figure}
\centering

\subfigure[ ]
{\begin{minipage}[b]{0.35\textwidth}
\includegraphics[width=1\textwidth]{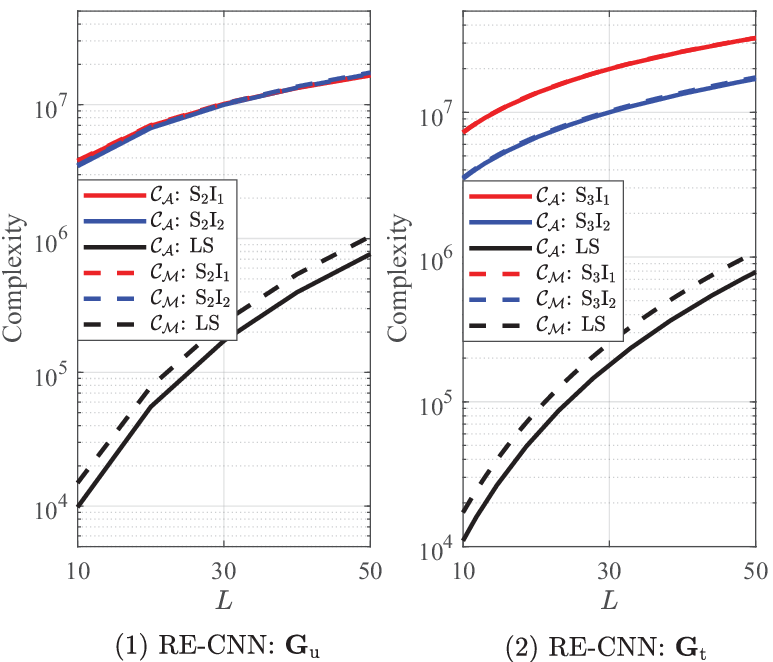}
\end{minipage}}
\subfigure[ ]
{\begin{minipage}[b]{0.35\textwidth}
\includegraphics[width=1\textwidth]{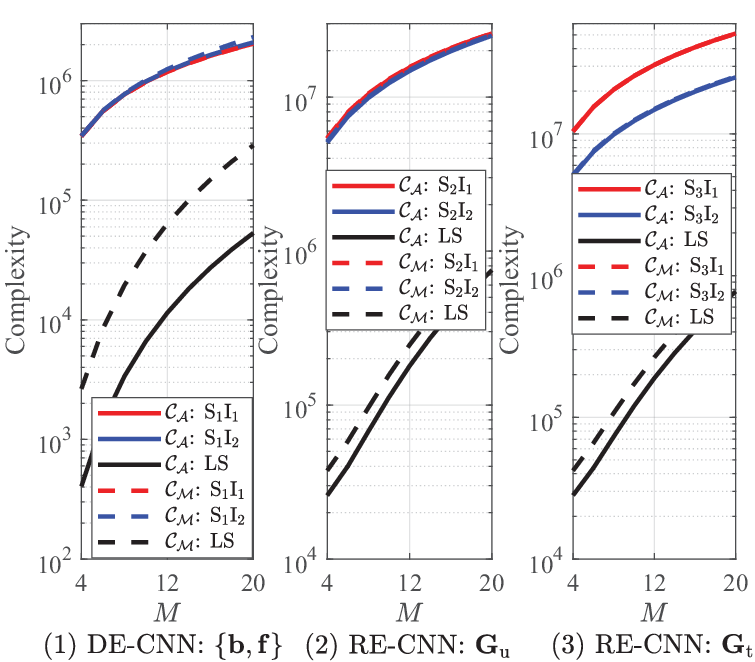}
\end{minipage}}


\caption{Complexity comparison for different channel dimension: (a) Complexity versus $L$ at $M=4$, (b) Complexity versus $M$ at $L=15$.} 
\label{fig:Complexity}
\end{figure}

The complexity of the proposed {DL} approach and \textcolor{black}{LS} baseline scheme is assessed in this subsection, considering the impact of increasing $L$ and $M$.
Based on the deduced formulations in Section \ref{sec:complexity}, the complexity of the proposed \textcolor{black}{DL} approach is measured by adding up the cost of the inputs generation and CNN-based online estimation.
The required number of real additions and multiplications, $\calC_\calA$ and $\calC_\calM$, is respectively shown in the solid and dashed curves in Fig. \ref{fig:Complexity}.

As depicted in Fig. \ref{fig:Complexity}, for the estimation of $\{\bfb,\bff\}$ and $\bfG_\rmu$, the complexity of the proposed \textcolor{black}{DL} approach trained by the first and second types of input-output pairs is comparable \textcolor{black}{to the LS baseline estimator}.
\textcolor{black}{For estimating} $\bfG_\rmt$, the proposed \textcolor{black}{DL} approach using $\big(\bfT^{\rmS_3\rmI_2}, \bfO^{\rmS_3}\big)$ attains lower complexity than that with $\big(\bfT^{\rmS_3\rmI_1}, \bfO^{\rmS_3}\big)$.
One can also observe from Fig. \ref{fig:Complexity} that the complexity of the proposed \textcolor{black}{DL} approach increases as $L$ or $M$ increases, and is higher than that of the \textcolor{black}{LS} baseline scheme.
It can be inferred from Figs. \ref{fig:NMSE_SNR}-\ref{fig:Complexity} that the proposed \textcolor{black}{DL} approach provides outstanding NMSE performance \textcolor{black}{at the expense of an} acceptable \textcolor{black}{computational} complexity \textcolor{black}{compared to the LS baseline scheme}.

\section{Conclusion}\label{sec:conclusion}
This paper has proposed a DL-based three-stage channel estimation approach, for the first time, for the IRS-assisted ISAC MISO system.
{The proposed} DL framework realized by the DE-CNN and RE-CNN has been developed to successively estimate the direct {SAC} channels, reflected communication channel, and reflected sensing channel.
{The pilot transmission protocol for the ISAC BS, UL UE, and IRS has been devised to facilitate the {SAC} channels estimation in each stage.}
Two types of input-output pairs have been designed for the CNNs.
Numerical results have indicated that under different SNR conditions, the proposed {DL} approach {provides} a considerable generalization capacity and significantly {enhanced} NMSE performance (e.g., $12.5\,\rm dB$ SNR improvement at $\text{NMSE}=10^{-1}$ for sensing channel estimation), compared to the \textcolor{black}{LS} baseline scheme.
Furthermore, under a wide range of channel dimensions, the proposed {DL} approach {possesses} a remarkable NMSE performance {improvement} over the {LS} baseline scheme.
The superiority of the proposed {DL} approach {is} attributed to the unique advantages of the designed input-output pairs, as well as the feature extraction and error correction abilities of the {proposed CNN architectures}.
The computational complexity of the proposed {DL} approach has been investigated, and \textcolor{black}{results showed} an acceptable complexity to achieve an accurate channel estimation.


\bibliographystyle{IEEEtran}

\bibliography{Reference_ISAC}

\begin{IEEEbiography}[{\includegraphics[width=1in,height=1.25in,clip,keepaspectratio]{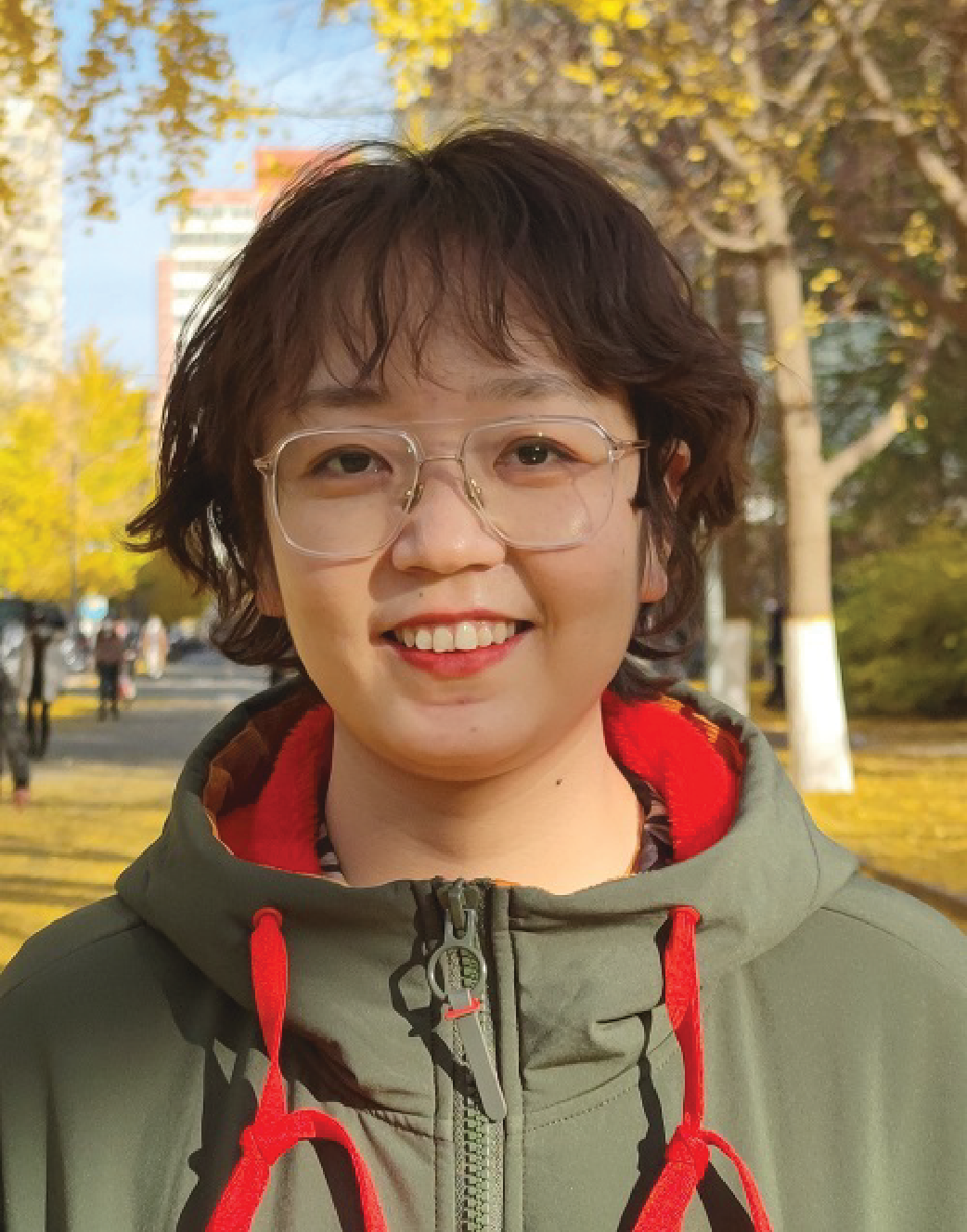}}]{Yu Liu} received the B.Eng. degree from the School of Electronic and Information Engineering, Beijing Jiaotong University, Beijing, China, in 2017. She is currently pursuing the Ph.D. degree with the State Key Laboratory of Rail Traffic Control and Safety, Beijing Jiaotong University, Beijing, China. Her current research interests include signal processing, interference cancellation, and integrated sensing and communication.
\end{IEEEbiography}

\begin{IEEEbiography}[{\includegraphics[width=1in,height=1.25in,clip,keepaspectratio]{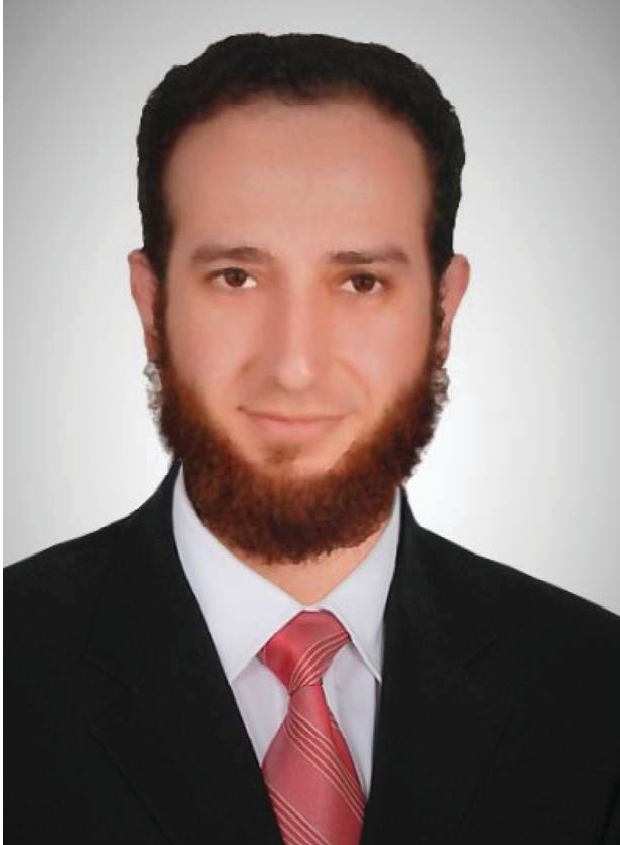}}]{Ibrahim Al-Nahhal} (M'18-SM'21) received the B.Sc. (Honours), M.Sc., and Ph.D. degrees in Electronics and Communications Engineering from Al-Azhar University in Cairo, Egypt-Japan University for Science and Technology, Egypt, Memorial University, Canada, in 2007, 2014, and 2020, respectively. From 2021 till now he works as a research associate and per-course instructor in Memorial University, Canada. Between 2008 and 2012, he was an engineer in industry, and a Teaching Assistant at the Faculty of Engineering, Al-Azhar University in Cairo, Egypt. From 2014 to 2015, he was a physical layer expert at Nokia (formerly Alcatel-Lucent), Belgium. He holds three patents. He serves as Editor of IEEE Wireless Communications Letters. He served as a Technical Program Committee and Reviewer for various prestigious journals and conferences. He was awarded the Exemplary Reviewer of IEEE Communications Letters in 2017. His research interests are reconfigurable intelligent surfaces, full-duplex communications, integrated sensing and communication, channel estimation, machine learning, design of low-complexity receivers for emerging technologies, spatial modulation, multiple-input multiple-output communications, sparse code multiple access, and optical communications.
\end{IEEEbiography}

\begin{IEEEbiography}[{\includegraphics[width=1in,height=1.25in,clip,keepaspectratio]{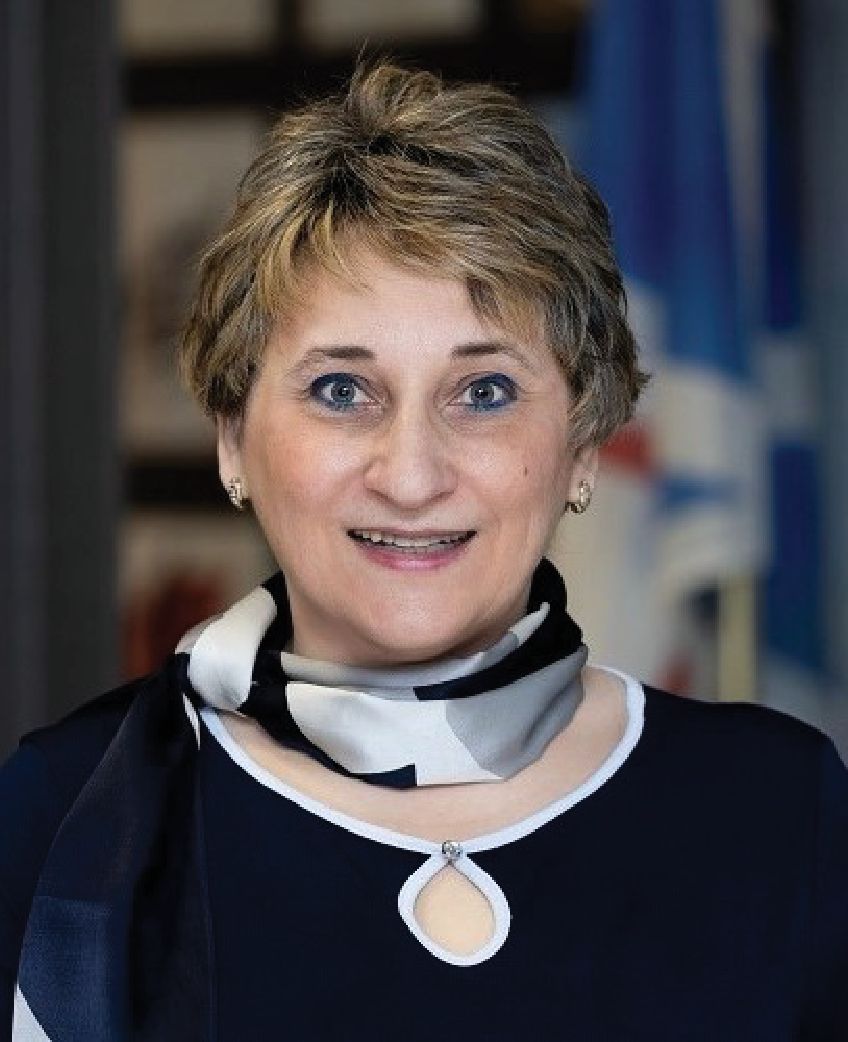}}]{Octavia A. Dobre} (Fellow, IEEE) received the Dipl. Ing. and Ph.D. degrees from the Polytechnic Institute of Bucharest, Romania, in 1991 and 2000, respectively. Between 2002 and 2005, she was with the New Jersey Institute of Technology, USA. In 2005, she joined Memorial University, Canada, where she is currently a Professor and Research Chair. She was a Visiting Professor with Massachusetts Institute of Technology, USA and Université de Bretagne Occidentale, France. Her research interests encompass wireless communication and networking technologies, as well as optical and underwater communications. She has (co-)authored over 400 refereed papers in these areas. Dr. Dobre serves as the Director of Journals of the Communications Society. She was the inaugural Editor-in-Chief (EiC) of the IEEE Open Journal of the Communications Society and the EiC of the IEEE Communications Letters. She also served as General Chair, Technical Program Co-Chair, Tutorial Co-Chair, and Technical Co-Chair of symposia at numerous conferences. Dr. Dobre was a Fulbright Scholar, Royal Society Scholar, and Distinguished Lecturer of the IEEE Communications Society. She obtained Best Paper Awards at various conferences, including IEEE ICC, IEEE Globecom, IEEE WCNC, and IEEE PIMRC. Dr. Dobre is an elected member of the European Academy of Sciences and Arts, a Fellow of the Engineering Institute of Canada, and a Fellow of the Canadian Academy of Engineering.
\end{IEEEbiography}

\begin{IEEEbiography}[{\includegraphics[width=1in,height=1.25in,clip,keepaspectratio]{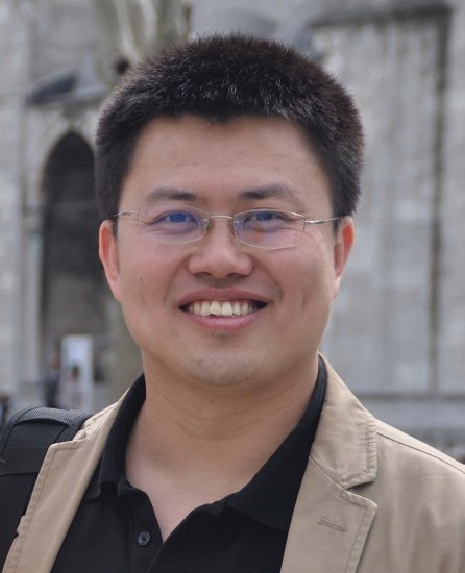}}]{Fanggang Wang} (S'10-M'11-SM'16) received the B.Eng. and Ph.D. degrees from the School of Information and Communication Engineering, Beijing University of Posts and Telecommunications, Beijing, China, in 2005 and 2010, respectively. He was a Post-Doctoral Fellow with the Institute of Network Coding, The Chinese University of Hong Kong, Hong Kong, from 2010 to 2012. He was a Visiting Scholar with the Massachusetts Institute of Technology from 2015 to 2016 and the Singapore University of Technology and Design in 2014. He is currently a Professor with the State Key Laboratory of Rail Traffic Control and Safety, School of Electronic and Information Engineering, Beijing Jiaotong University. His research interests are in wireless communications, signal processing, and information theory. He served as an Editor for the IEEE COMMUNICATIONS LETTERS and technical program committee member for several conferences.
\end{IEEEbiography}

\end{document}